\documentclass{SciPost}

\hypersetup{
    colorlinks,
    linkcolor={red!50!black},
    citecolor={blue!50!black},
    urlcolor={blue!80!black}
}

\newcommand{\gev}{\text{GeV}}

\newcommand{\mgfive}{\textsc{MG5\_aMC@NLO}}

\usepackage{easyReview}
\usepackage{cleveref}

\usepackage[compat=1.0.0]{tikz-feynman}
\usepackage[bitstream-charter]{mathdesign}
\usepackage[normalem]{ulem}
\usepackage{multirow}
\usepackage{graphicx}
\usepackage{subcaption}
\usepackage{caption}
\usepackage{booktabs} 

\DeclareSymbolFont{usualmathcal}{OMS}{cmsy}{m}{n}
\DeclareSymbolFontAlphabet{\mathcal}{usualmathcal}

\fancypagestyle{SPstyle}{
\fancyhf{}
\lhead{\colorbox{scipostblue}{\bf \color{white} ~SciPost Physics }}
\rhead{{\bf \color{scipostdeepblue} ~Submission }}

\fancyfoot[C]{\textbf{\thepage}}
}

\begin{document}

\pagestyle{SPstyle}

\begin{center}{\Large \textbf{\color{scipostdeepblue}{
Precision physics at the muon collider: $m_W$ and CKM matrix elements
}}}\end{center}

\begin{center}\textbf{
Ludovica Aperio Bella\textsuperscript{1$\star$} 
Roberto Franceschini\textsuperscript{2,3$\dagger$} 
Federico Meloni\textsuperscript{1$\ast$} 
Xing Wang\textsuperscript{2$\div$} 
}\end{center}

\begin{center}
{\bf 1} Deutsches Elektronen-Synchrotron DESY, Notkestrasse 85, 22607 Hamburg, Germany\\
{\bf 2} Universit\`a degli Studi Roma Tre and INFN, Via della Vasca Navale 84, 00146 Rome, Italy\\
{\bf 3} Theoretical Physics Department, CERN, 1211 Geneva 23, Switzerland
\\[\baselineskip]
$\star$ \href{mailto:email1}{\small ludovica.aperio.bella@desy.de}\,,
$\dagger$ \href{mailto:email2}{\small roberto.franceschini@uniroma3.it}\,,
$\ast$ \href{mailto:email2}{\small federico.meloni@desy.de}\,,
$\div$ \href{mailto:email1}{\small xing.wang@uniroma3.it}
\end{center}

{\tiny{\hfill  CERN-TH-2026-092, DESY-26-068}}
\section*{\color{scipostdeepblue}{Abstract}}
\textbf{\boldmath{%
We examine the potential for a 10~TeV lepton collider to carry out precision measurements of the W boson mass and W boson couplings strength, i.e. the CKM matrix elements. We consider the several W boson production mechanisms and focus on the most copious at 10~TeV, that is effective $\gamma W \to W$, a process viable at both opposite sign and same-sign leptonic colliders.
We find that  the leptonic W decay channel can hardly be competitive with present determinations, due to lack of rate.  
The hadronic channel has potential to improve over the current $\simeq$10~MeV from measurements at hadron colliders, motivating detector developments towards high-precision hadronic energy measurements.
We find that the precision understanding of  the detector response to hadrons can also lead to a determination of the CKM matrix elements. We expect determination of CKM matrix elements surpassing by far the present precision for couplings involving heavy quarks, notably $V_{cb}$, avoiding the present bottle-necks due to poor knowledge of hadronic matrix elements needed in low energy extractions of CKM matrix elements.
Our findings motivate detector developments towards high-precision hadronic energy measurements and flavor tagging.
}}

\vspace{\baselineskip}

\noindent\textcolor{white!90!black}{%
\fbox{\parbox{0.975\linewidth}{%
\textcolor{white!40!black}{\begin{tabular}{lr}%
  \begin{minipage}{0.6\textwidth}%
    {\small Copyright attribution to authors. \newline
    This work is a submission to SciPost Physics. \newline
    License information to appear upon publication. \newline
    Publication information to appear upon publication.}
  \end{minipage} & \begin{minipage}{0.4\textwidth}
    {\small Received Date \newline Accepted Date \newline Published Date}%
  \end{minipage}
\end{tabular}}
}}
}


\vspace{10pt}
\noindent\rule{\textwidth}{1pt}
\tableofcontents
\noindent\rule{\textwidth}{1pt}
\vspace{10pt}


\section{Introduction \label{sec:intro}}

The physics case for a muon collider is very strong, as it has several unique opportunities to probe new aspects of Standard Model physics as well as to search for new physics beyond the Standard Model. 
One aspect of muon collider physics that has so far received a smaller share of the attention is the possibility to have a precision physics programme. Such a programme aims to improve upon measurements of key properties like particle masses and couplings performed at previous machines.  Natural candidates to carry out such a programme are the $t$ quark~\cite{Franceschini:2022veh}, $W$ and $Z$ bosons whose couplings and masses are key properties of the SM and can be used to probe new physics in a number of ways.

In this work we deal with the possibility to measure the $W$ boson mass and the overall strength of its coupling, i.e. the CKM matrix elements. These will hinge upon detector performance in areas such as kinematic reconstruction and flavor tagging, which are also key to carry out the ``Higgs factory'' part of the muon collider physics programme. 

The result we obtain for $W$ bosons demand to push the knowledge of the detector to its extremes, aiming for relative precision on the mass scale at well below $10^{-3}$ level and similarly demanding targets for flavor tagging. In this sense the goal posts set by these measurements can be conducive to improvements in theory and detector development  for the muon collider. In addition, these measurements could be the leading ones at the epoch in which the muon collider will be operating.
Much depends on the possibility that the experiments at the HL-LHC will measure $m_W$ better than the current 10~MeV level~\cite{CMS:2024lrd} or find  ways to access the CKM matrix element with precision better than or comparable with low energy determinations, e.g. from meson decays.
In any event, the method and the understanding of the detector that is required to carry out the measurements that we propose are going to be useful at the very least as calibration, a precious foundation to build upon for the rest of the muon collider physics programme. 

For the measurements of CKM matrix elements that we propose here we stress that they are free from issues that plague low-energy flavor experiments. In particular the high $q^2\gg m^2_q$ that is characteristic of the decay $W\to q q^
\prime$ renders these measurements much less dependent on the systematic uncertainty from non-perturbative physics, which currently are a bottleneck for low $q^2$ flavor experiments, with no prospect to do better than 1\% in the foreseeable future~\cite{ATLAS:2025lrr}.

In our study we consider several possible $W$ boson production mechanisms at lepton colliders, namely pair production $\ell \ell \to WW + X $ and single $\ell \ell \to W +X $ which contain

\begin{eqnarray}
  \ell^+ \ell^- &\to& W^+W^-  \label{eq:annihilation-pair} \,, \\ 
\ell^+ \ell^- &\to& W^+W^- \nu \nu  \label{eq:vbf-pair}\,, \\ 
\ell^+ \ell^- &\to& W^+W^- \ell^+ \ell^-  \label{eq:vbf-neutral-pair}\,, \\ 
\ell^+ \ell^- &\to& W^\pm  \nu \ell^\mp \,. \label{eq:effectivephoton-single}
\end{eqnarray}

The total cross-sections for these rates are tabulated at 3, 10, and 30~TeV in  \Cref{tab:rates} using the Effective Photon Approximation for eq.~\eqref{eq:vbf-neutral-pair} and eq.~\eqref{eq:effectivephoton-single}.
\begin{table}[h!]
    \centering
    \begin{tabular}{c|ccc}
                                      & 3~TeV & 10~TeV & 30~TeV \\
                                      \hline
     $\ell^+ \ell^- \to W^+W^-$       &  0.45~pb & 57~fb& 8~fb\\
      $\ell^+ \ell^- \to W^+W^- \nu \bar{\nu}$     & 0.12~pb & 0.40~pb & 0.83~pb\\
      $\mu^+ \mu^- \to W^+W^- \mu^+ \mu^- $     & 1.2~pb & 3.5~pb & 7.2~pb\\
      $e^+ e^- \to W^+W^- e^+ e^- $     & 3.3~pb & 8.9~pb & 17~pb\\
   $\mu^+ \mu^- \to W^\pm  \nu\ \mu^\mp$ & 17.8~pb & 31~pb &  47~pb \\
   $e^+ e^- \to W^\pm  \nu\ e^\mp$ & 28~pb & 48~pb &   69~pb \\
      $\ell^+ \ell^+ \to W^+W^+ \bar{\nu} \bar{\nu}$     & 0.13~pb & 0.43~pb & 0.86~pb\\
      $\mu^+ \mu^+ \to W^+W^- \mu^+ \mu^+ $     & 1.2~pb & 3.5~pb & 7.2~pb\\
      $e^+ e^+ \to W^+W^- e^+ e^+ $     & 3.3~pb & 8.9~pb & 17~pb\\
   $\mu^+ \mu^+ \to W^+  \bar{\nu}\ \mu^+$ & 8.9~pb & 15~pb &  23~pb \\
   $e^+ e^+ \to W^+  \bar{\nu}\ e^+$ & 14~pb & 24~pb &   34~pb \\
   \end{tabular}

  \caption{\label{tab:rates}  Rates for the processes eqs.~(\ref{eq:annihilation-pair})-(\ref{eq:effectivephoton-single}) at 3, 10, and 30~TeV. When the rate does not depend on the initial state being electron or muon a collective result for $\ell$ is given, otherwise the lepton flavor is specified explicitly. Rates are fully inclusive and are computed at LO with massive leptons using \mgfive \cite{Alwall:2014hca}}
\end{table}

\begin{figure}[h!]
  \centering
  \begin{subfigure}{0.22\linewidth}
    \includegraphics[width=\linewidth]{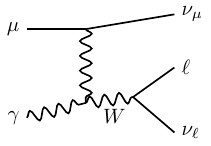}
    \caption{}   
  \end{subfigure}
  \begin{subfigure}{0.22\linewidth}
    \includegraphics[width=\linewidth]{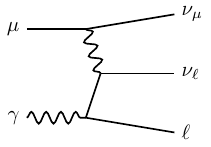}
    \caption{}   
  \end{subfigure}
  \begin{subfigure}{0.22\linewidth}
    \includegraphics[width=\linewidth]{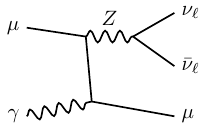}
    \caption{}   
  \end{subfigure}
  \begin{subfigure}{0.20\linewidth}
    \includegraphics[width=\linewidth]{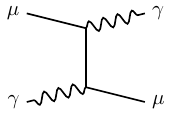}
    \caption{}   
  \end{subfigure}
  \caption{ Representative Feynman diagrams for the W production (a), non-resonant background (b), background from on-shell $Z$ decay (c), and background from Compton scattering (d). }
  \label{fig:feyn}
\end{figure}

The most abundant process for 10~TeV is Eq.~\eqref{eq:effectivephoton-single}, which is dominated by the effective collision of $\gamma$ on $\mu^\pm$
\begin{eqnarray}
  \label{eq:gammaWtoW} \gamma \ell^\pm \to W^\pm \nu  \,.
\end{eqnarray}
The full signal process is drawn in the leftmost diagram of \Cref{fig:feyn}.
Owing to $\gamma$ being radiated equally well by $\mu^+$ and $\mu^-$, this process is viable both at $\mu^+\mu^-$\cite{InternationalMuonCollider:2025sys} and $\mu^+\mu^+$\cite{Hamada:2022mua} machines, as well at same energy $e^+e^-$ colliders~\cite{Chigusa:2025azs}, though with not identical cross-section e.g. due to different logarithmic enhancements of the photon luminosity for $\mu$ and $e$ beams. Up to these small differences our analyses can be applied to both type of machines.

At a different energy than 10~TeV a different analysis should be performed picking the relevant most abundant channel for $W$ boson production at each machine. In this work we consider only 10~TeV and leave other possibilities for future work. 
In \Cref{sec:detector} we describe the detector that we imagine to operate at the 10~TeV $\mu^+\mu^-$ machine. In \Cref{sec:mW-leptonic} we describe the sensitivity to $m_W$ in the leptonic decay channel, following the recent measurements of ATLAS, CDF and CMS~\cite{1307.7627v2,Kotwal:2025xsy,CDF:2022hxs, ATLAS-CONF-2023-004,CMS:2024lrd,ATLAS-Collaboration:2017ac}. In \Cref{sec:Wmass-hadronic} we discuss the use of hadronic final states and highly precise calibration of the detector to carry out a $W$ boson mass measurement. In \Cref{sec:CKM} we describe the reach for precision measurements of CKM matrix elements. Finally in \Cref{sec:conclusion} we give our conclusions.

\section{Detector description \label{sec:detector}}

Two distinct detector concepts were developed to take data at a $\sqrt{s}=10$~TeV muon collider: MUSIC~\cite{Andreetto:2025mrd} (MUon System for Interesting Collisions) and MAIA~\cite{MAIA:2025hzm} (Muon Accelerator Instrumented Apparatus). 
Both detector designs have been optimised to minimise the impact of the machine-induced backgrounds while at the same time maintaining high efficiency and accuracy for physics measurements exploiting both low- and high-momentum reconstructed objects. They share a similar structure, conventional of multipurpose collider experiments, with a cylindrical layout hosting several purpose-specific subdetectors and a solenoidal magnet to generate an axial magnetic field. While the two detector concepts differ in the specific technologies that were selected for some of the sub-detectors and in details of the layout, they are generally expected to deliver comparable performances.

In the following, the MAIA detector concept was selected to carry out our detailed studies. The MAIA detector is designed to be a multi-purpose particle detector with cylindrical geometry and a near 4$\pi$ coverage in solid angle\footnote{A right-handed reference system is used with the origin at the center of the detector, the nominal collision point: the z-axis is aligned with the direction of the clockwise-circulating $\mu^{+}$ beam, the y axis points upward, and the x axis lies on the plane of the collider ring.}. It consists of an inner tracking system surrounded by a superconducting solenoid providing a 5~T axial magnetic field, electromagnetic and hadronic calorimeters, and a muon spectrometer. 
The inner tracking system covers the range $ 10^\circ < \theta < 170^\circ$. It consists of two silicon subdetectors: the vertex detector, a high-resolution pixel detector closest to the collision point and the inner tracker, a macro-pixel detector. A high-granularity silicon-tungsten sampling calorimeter is dedicated to electromagnetic energy measurements, while an iron-scintillator sampling calorimeter provides measurements for hadrons. An external muon system, based on RPC chambers, provides muon particle identification capabilities. The number of secondary particles from beam-induced backgrounds entering the detector is suppressed by placing massive nozzle-shaped absorbers in close proximity of the interaction point, which define the inner detector envelope and the angular acceptance of the detector. 

\section{Leptonic channel \label{sec:mW-leptonic}}

The cleanest way to observe a $W$ boson is to search for its leptonic decay, as it is indeed the case for the most precise measurements to date \cite{1307.7627v2,Kotwal:2025xsy,CDF:2022hxs, ATLAS-CONF-2023-004,CMS:2024lrd,ATLAS-Collaboration:2017ac}. This choice is dictated by the favorable signal-to-noise ratio that this choice entails at hadron machines. As we will discuss in the following, this is not a compulsory choice at leptonic machines, and it might be more profitable to pursue hadronic decays. 
Nevertheless, this clean channel is useful for our analysis to show the precision that can be attained using leptons in the final state of muon collisions and to assess the possible utility of the $W$ boson as a source for calibration of the detectors in the range of energy and transverse momentum populated by $W$ boson production.

\subsection{Lab-frame analysis}

For the leptonic channel,
\begin{equation}
    \mu^+\mu^-\rightarrow W^\pm \mu^\mp\nu_\mu \rightarrow \ell^\pm \mu^\mp\nu_\ell\nu_\mu, \label{eq:signal}
\end{equation}
the energy of the outgoing charged leptons in the rest frame of the $W$ boson is mono-chromatic $E^*_\ell = m_W/2$, where $\ell$ indicates the leptons from the decay of the $W$ boson.  However, because of the two undetectable neutrinos, the rest frame of the $W$ boson cannot be fully reconstructed, and the energy distribution of the charge leptons in the lab frame gets smeared by the boost of the $W$ bosons. However, such energy distribution in the lab frame can still be used to extract the $W$ boson mass. In this section, we perform a template-fitting to the parton level energy distribution of the charged lepton in the lab frame to extract the $W$ boson mass.

\subsubsection{Definition of the signal and event selection}
For the signal \cref{eq:signal}
we simulate the parton level events using the effective photon approximation~(EPA)~\cite{vonWeizsacker:1934nji, Williams:1934ad, Frixione:1993yw},
\begin{equation}
    \mu^\pm\gamma\rightarrow W^\pm \nu_\mu \rightarrow \ell^\pm\nu_\ell\nu_\mu, \label{eq:signal-epa}
\end{equation}
with \mgfive~\cite{Alwall:2014hca}.
Events are   selected with transverse momentum for the final state leptons 
\begin{equation}
    p_{T,\,\ell^\pm} > 20~{\rm GeV},
\end{equation}
and angular acceptance
\begin{equation}
    10^\circ<\theta_{\ell}<170^\circ.
    \label{eq:cut-angular}
\end{equation}
The latter angular requirement is due to the geometrical acceptance of the detectors at the muon collider, which have a non-instrumented region around the beam pipe to shield the active detector volume from noise due to beam activity.

As we use the lepton energy distribution to fit the $W$ boson mass, it is crucial to achieve great precision on the lepton energy measurement. 
In the following we consider  $\ell=e,\mu$ and  we discuss the possible difference in precision between muon and electron reconstruction due to different backgrounds, e.g. the muon channel pollution from Bhabha scattering. The energy and transverse momentum distributions of the muon for a representative $m_W$ are shown in \Cref{fig:dist_mw} in black lines. As expected, the distributions peak around $m_W/2$. The precise dependence of these spectra on $m_W$ will be templated and will enable the extraction of $m_W$ from the observation of the measured  spectra.

\subsubsection{Backgrounds}

\begin{figure}[tb]
    \centering
    \includegraphics[width=0.45\linewidth]{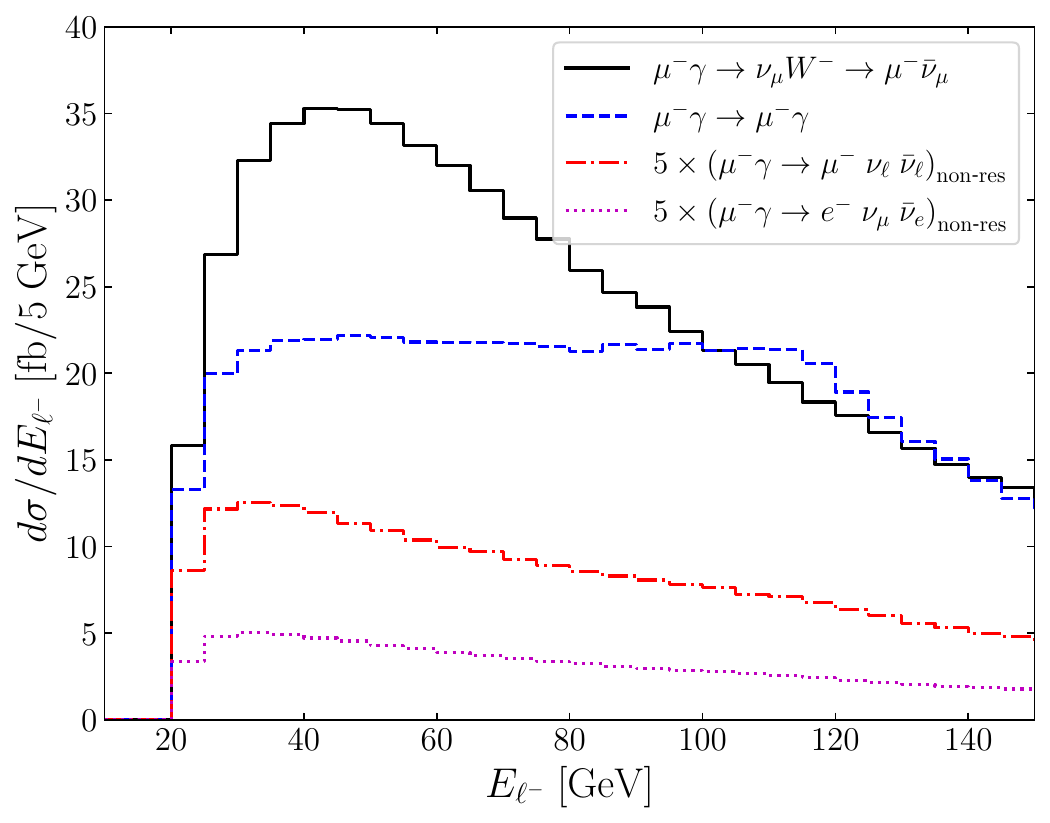}
    \includegraphics[width=0.45\linewidth]{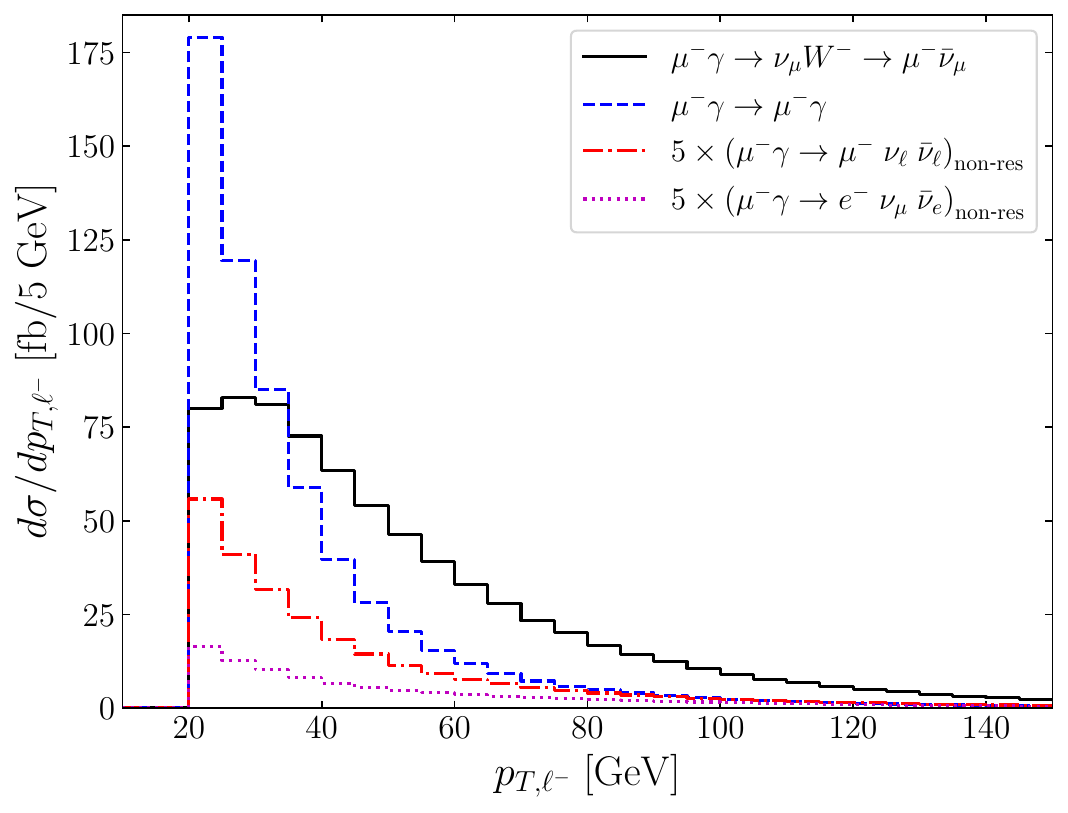}
    \caption{The energy (left) and transverse momentum (right) distributions of the final state muons, simulated at the parton level for $\sqrt{s} = 10$~TeV.}
    \label{fig:dist_mw}
\end{figure}

In the electron channel, the dominant background is given by the irreducible non-resonant production
\begin{equation}
    \mu^\pm \gamma\rightarrow e^\pm\,\nu_e\,\nu_\mu, \label{eq:background-non-resonant}
\end{equation} 
where we have used the EPA, hence we assume that the missing collinear muon carries negligible $p_T$. We simulate all the parton level backgrounds
using \mgfive~v3.5.6~\cite{Alwall:2014hca}. The energy and transverse momentum distributions  of the electron for such background are shown in \Cref{fig:dist_mw} in purple dotted lines, multiplied by a factor of 5 to ease the visualization. We note that the energy distribution is smooth and monotonically decreasing near the signal peak $E_\ell \sim m_W/2$. The shape of the main background is rather flat around the peak region, thus creating an ideal situation for the extraction of the peak. For these reasons in the following we choose to use the lepton energy distribution to extract $m_W$.

While $\mathcal{O}(40~{\rm GeV})$ muons typically have much better energy uncertainty than electrons, the muon channel receives much larger background. The irreducible non-resonant background,
\begin{equation}
    \mu^\pm\gamma\rightarrow \mu^\pm\,\nu_\ell\,\nu_\ell, \label{eq:background-non-resonant-mu}
\end{equation} 
becomes larger since it now also contains neutrino pairs from on-shell $Z$ decays,
\begin{equation}
    \mu^\pm\gamma\rightarrow \mu^\pm Z\rightarrow \mu^\pm \left(\nu_\ell\,\bar{\nu}_\ell\right),
\end{equation} 
where indices $\ell = e,\, \mu,\, \tau$. The spectrum from this background is shown in \Cref{fig:dist_mw} in red dotted-dashed lines, multiplied by a factor of 5 to ease the visualization. Another more dominant background in the muon channel is an effective Compton scattering with EPA photons,
\begin{equation}
    \mu^\pm\gamma\rightarrow \mu^\pm\gamma, \label{eq:background-bhabha}
\end{equation}
where the final state photon falls outside the detector acceptance, which is missed for any of the following reasons:
\begin{equation}
    E_\gamma < 10~{\rm GeV},\quad {\rm or}\quad \theta_\gamma < 10^\circ,\quad {\rm or}\quad \theta_\gamma > 170^\circ.
\end{equation}
Such process has very high rate owing to the large cross-section of the Bhabha scattering, as shown in  \Cref{fig:dist_mw} in blue dashed lines. 

\begin{figure}[t!]
  \centering
  \includegraphics[width=0.45\linewidth]{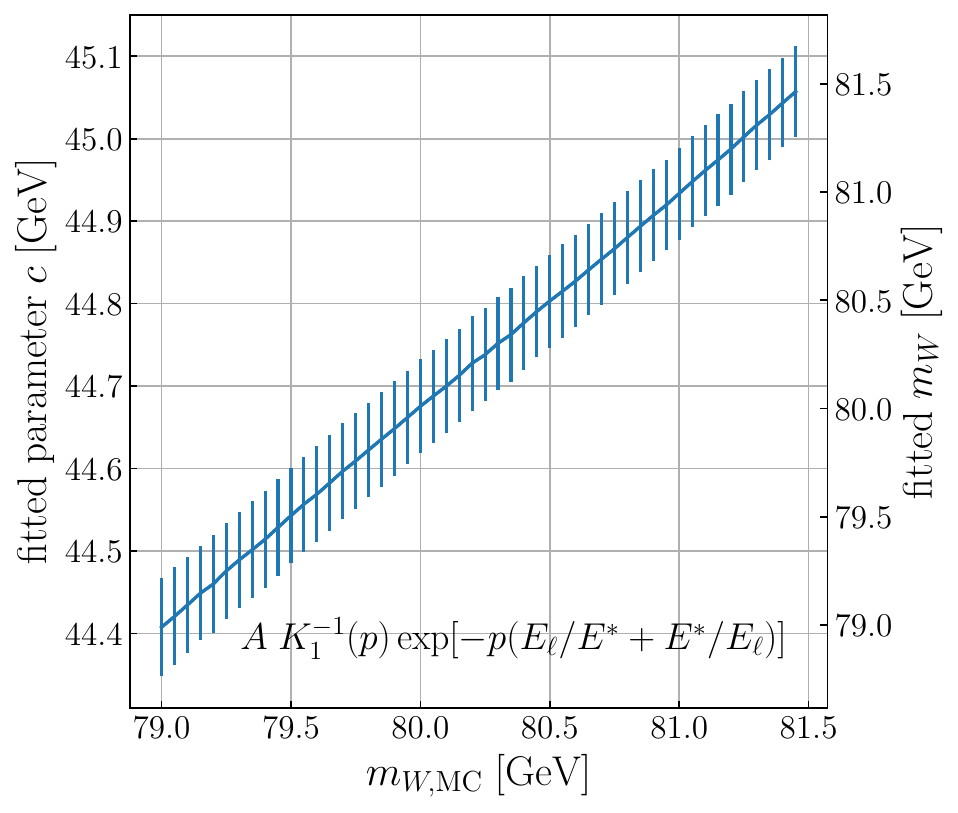}
  \includegraphics[width=0.4\linewidth]{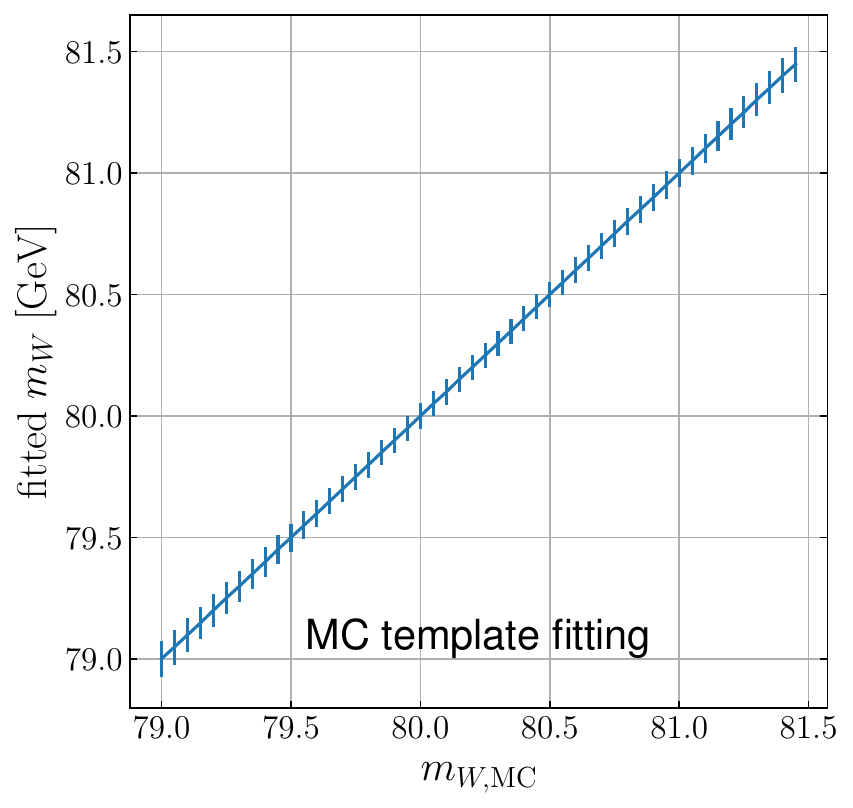}
  \caption{ (left) The parameter $E^*$ obtained from the fit of the template \cref{eq:bessel-template} versus the truth level $m_W$. (right) The best fit $m_W$ from a fully numerical template fitting against the truth level $m_W$. Both panels are for the muon channel. }
  \label{fig:fitting_mw}
\end{figure}

\subsubsection{Analytical templates}
For the extraction of $m_W$ from the spectrum we explore two possible routes. One is based on giving more importance to the peak region of the energy spectrum, which has been shown in Refs.~\cite{Agashe:2012bn,Agashe:2015ike} to be less prone to corrections from higher orders in perturbation theory in modeling the production of resonances.
Motivated by the findings of these works we choose to fit the lepton energy distribution with the ansatz,
\begin{equation}
    \dfrac{dN}{dE_\ell} \sim A\cdot \exp\left[-p\left(\dfrac{E_\ell}{E^*} + \dfrac{E^*}{E_\ell}\right)\right]. \label{eq:bessel-template}
\end{equation}
The parameter $E^*$, corresponding to the maximum point of the distribution, contains the information of the $W$ boson mass. The other parameters $A$ and $p$ are less sensitive to $m_W$. Thus, we can easily diagnose the results from this method and anticipate what precision can be attained for a given luminosity. 

For our study we divide the energy distribution into 44 bins between $30-90$ GeV, and fit it to the analytic ansatz. The left panel in Fig.~\ref{fig:fitting_mw} shows the correlation between the fitted parameter and the truth $m_{W,{\rm MC}}$ in the MC simulation, with the error bars indicating the $1\sigma$ fit uncertainties at 10~TeV muon collider with $10~{\rm fb}^{-1}$ luminosity.

\subsubsection{Numerical templates fitting}
While the fitting of an analytical model lends itself to easier interpretation and diagnose of the results, it leaves some unexploited potential for the extraction of $m_W$. For instance the normalization parameter $A$ and the shape parameter $p$ in \cref{eq:bessel-template} contain some additional information on $m_W$, which is not  exploited in the left panel of \Cref{fig:fitting_mw}.

To maximize the amount of information on $m_W$ from the full shape of the energy spectrum we perform a fully numerical template-fitting study.
The lepton energy distribution is divided into 44 bins between $30-90$ GeV. We simulate signal events for 50 different $W$ boson masses between $79-81.5$ GeV. Then, for the number of events in each bin, we use linear interpolation in terms of $W$ boson mass, and perform a chi-square fit of the $W$ boson mass value whose template fits best with some pseudo-data that we generate for a representative $m_W$, that we denote as $m_{W,{\rm MC}}$.

The correlation between the fitted $m_W$ and the truth $m_{W,{\rm MC}}$ in the MC simulation is shown in the right panel in Fig.~\ref{fig:fitting_mw}, with the error bars indicating the $1\sigma$ fit uncertainties from the $\chi^2$ variation at a 10~TeV muon collider with $10~{\rm ab}^{-1}$ luminosity.

\begin{figure}[t!]
  \center
  \includegraphics[width=0.5\linewidth]{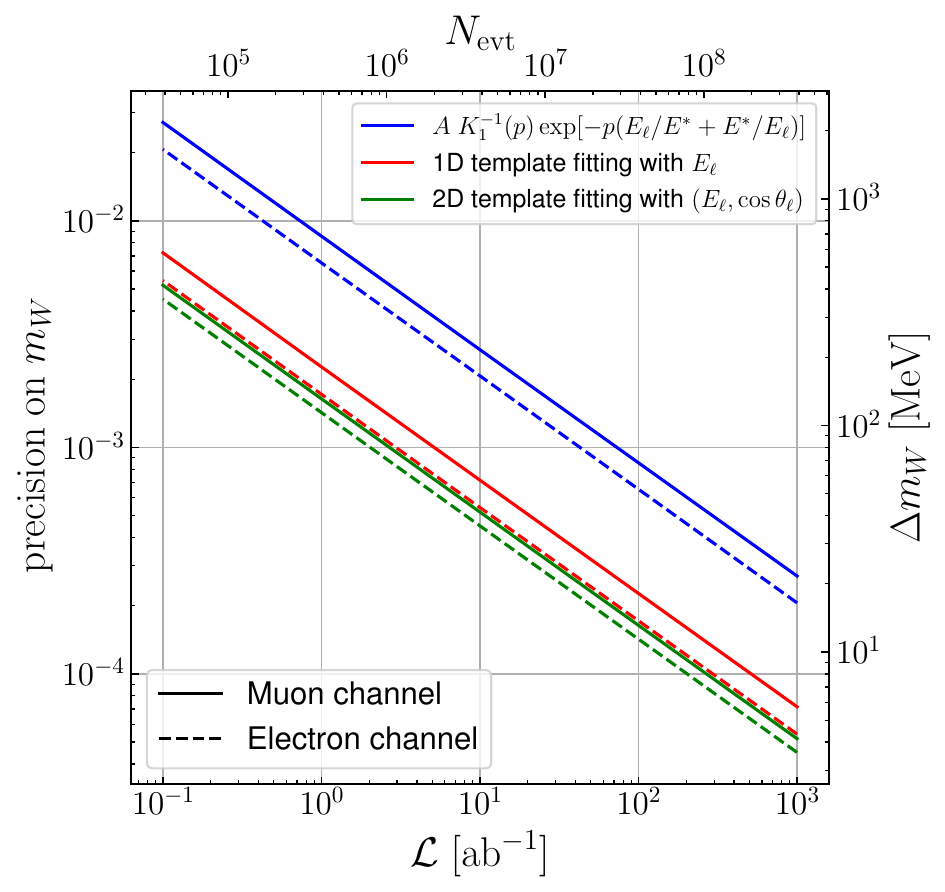}
  \caption{\label{fig:mWvsLumi-leptonic}Expected precision on $m_W$ as function of the luminosity.}
  \end{figure}

A comparison between the two fitting methods is demonstrated in Fig.~\ref{fig:mWvsLumi-leptonic}. The projected precision on $m_W$ for each fit strategy is plotted as a function of the collider luminosity. With template fitting, shown in red line, one can achieve the precision of $0.72\times 10^{-3}$ from the muon channel and $0.54\times 10^{-3}$ from the electron channel, with $10~{\rm ab}^{-1}$ luminosity. The fitting using the analytic ansatz parameter $E^*$ leads to a precision of $2.7\times 10^{-3}$ from the muon channel and $2.1\times 10^{-3}$ from the electron channel, with $10~{\rm ab}^{-1}$ luminosity, due to the unexploited sensitivity in other parameters in the fit and the distortion of the distribution caused by selection cuts and the background. The dashed lines in Fig.~\ref{fig:mWvsLumi-leptonic} shows the precision in the electron channel, which is somewhat better than the muon channel due to absence of the background \cref{eq:background-bhabha}.

\subsection{Beyond univariate lepton analysis}
\begin{figure}[t!]
  \center
  \includegraphics[width=0.5\linewidth]{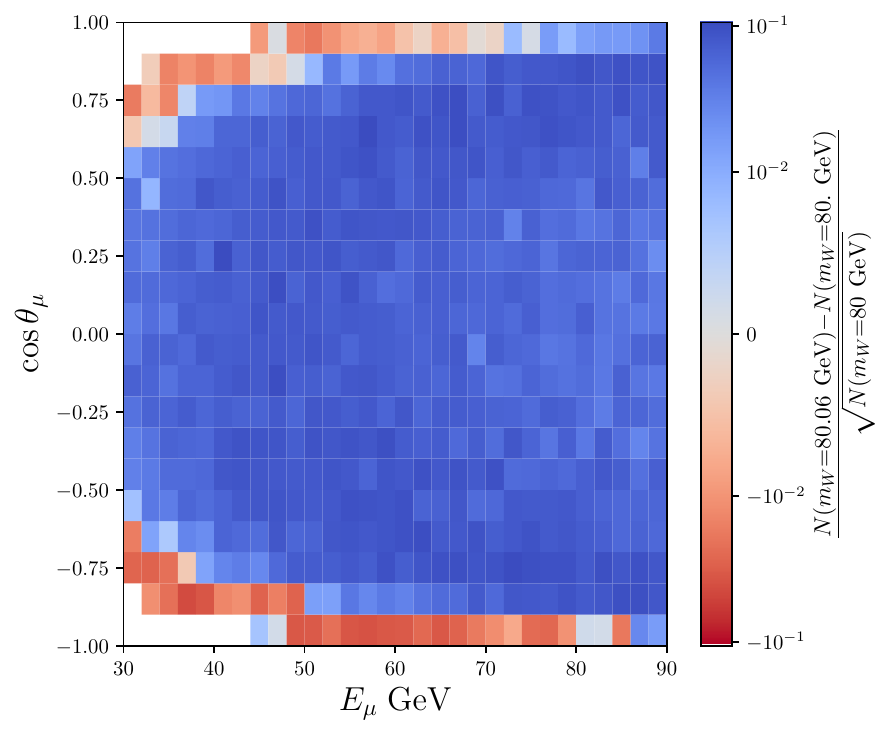}
  \caption{\label{fig:2d_heatmap} The difference in the numbers of events $N$ between $m_W = 80.06$~GeV and $m_W = 80$~GeV, in the 2D $E_\mu\text{-}\cos\theta_\mu$ distribution, relative to the statistical fluctuation $\sqrt{N}$, at $\sqrt{s} = 10$~TeV with $L=10~{\rm ab}^{-1}$ luminosity.}
  \end{figure}

The results from the study of 1D templates (analytical or numerical) demonstrates that the muon collider reach precision below $10^{-3}$ in the extraction of $m_W$. Such precision is not competitive with the present determination of $m_W$, thus it would likely be useful as a mean to constrain detector response, i.e. a standard candle process from which carry out calibration of the muon collider detector.

One of the main reasons the extraction does not reach the $10^{-4}$ level is that the information on the $W$ boson mass is spread over a wide rise-and-fall shape for the energy (or $p_T$) distribution. Increased sensitivity can be achieved by considering distributions with sharper features or using more information, e.g. by considering in full the differential information on the measured lepton.

Sharper features can be obtained, for instance, if on an event-by-event basis the energy of the lepton could be boosted in a frame in which the $W$ boson is at rest, or nearly at rest. This would recover, at least partly, the sharp $E_\ell=m_W/2$ feature of which the energy spectrum is in principle endowed. Research to gain this event-by-event information on the boost is left for future work.

In this subsection, we pursue the extraction of the $W$ boson mass through 2D template fitting in the $(E_\mu, \cos\theta_\mu)$ space. In the fitting, the 2D $E_\mu\text{-}\cos\theta_\mu$ distribution is divided into $30\times 20$ {equal-sized} bins in the range
\begin{equation}
    30~{\rm GeV} < E_\mu < 90~{\rm GeV},\qquad -1 < \cos\theta_\mu < 1.
\end{equation}
{The edge bins in $\cos\theta$ contain data only up to the detector acceptance $|\cos\theta|<0.985$, as for the 1D analysis. We formulate numerical templates in this 2D space and extract $m_W$ with strategy similar to what we have done in the 1D case.}

Figure~\ref{fig:2d_heatmap} shows the difference in the numbers of events $N$, in the 2D $E_\mu\text{-}\cos\theta_\mu$ distribution, for benchmarks with $m_W = 80.06$~GeV and $m_W = 80$~GeV, relative to the statistical fluctuation $\sqrt{N}$, at $\sqrt{s} = 10$~TeV with $L=10~{\rm ab}^{-1}$ luminosity. The lepton from the $W$ boson decay tends to be more central for heavier $W$ masses. The projected precisions on $m_W$ from the 2D template fitting is plotted in green lines in Fig.~\ref{fig:mWvsLumi-leptonic}. A precision of $0.52\times 10^{-3}$ from the muon channel or $0.45\times 10^{-3}$ from the electron channel can be achieved with $10~{\rm ab}^{-1}$ luminosity. As expected, the use of more differential information brings more sensitivity to $m_W$, thus this mass measurement can be thought as a useful process on which calibrate the detector with some degree of differentiation in the angular dimension.

\section{Hadronic $W$ decay channel\label{sec:Wmass-hadronic}}

As discussed already in early work on the International Linear Collider~\cite{Baak:2013fwa,Anguiano:2020qpk,Wilson2023}, it is envisageable to measure $m_W$ at high precision in leptonic collisions using hadronic $W$ boson decays. This channel enjoys the largest possible rate, but also faces the greatest difficulties as it is challenging to obtain a precise measurement of the energy of a hadronic jet and to understand the energy scale associated to hadronic activity. 
The subject of the calibration of hadronic jets is an active field of research~\cite{Podolanski:1954dyc,Rodriguez:2020qhf}.
To begin to explore the physics potential   of such measurement at the muon collider we carry out in the following a full detector simulation of hadronic $W$ boson decays and extraction of the $W$ boson mass. Significant additional work is needed to attain the sought precision at a muon collider.

\subsection{Description of the detector simulation}
\label{subsec:simreco}

The studies on the hadronic $W$ boson decays have been performed using a detector simulation based on Geant4~\cite{Geant4}. The detector simulation and event reconstruction were performed with the \textsc{MuonColliderSoft} software stack~\cite{Bartosik:2021bjh,Accettura:2023ked} within the Key4hep software ecosystem~\cite{FernandezDeclara:2022voh,Key4hep:2023nmr}. The detector geometry was implemented for simulation with the DD4hep detector description toolkit~\cite{dd4hep}, which also supplies the necessary interfaces to Geant4 to simulate the detector response via the \texttt{ddsim}~\cite{ddsim} tool. Beam backgrounds produced by the muon decays in the latest accelerator lattice were modelled with the FLUKA~\cite{BATTISTONI201510,FLUKA-new} particle transport code up to their entry in the detector volume. Similarly, incoherent pairs produced at the interaction point are simulated with the GUINEA-PIG~\cite{Schulte:1997nga} code, and transported with FLUKA up to the detector volume. Both beam-related backgrounds were simulated in the detector volume using the same software used to process the signal processes.

Each collision event is reconstructed from a collection of detector information using dedicated algorithms. Tracks are reconstructed in the tracking system within $|\eta| < 2.44$ using the ACTS~\cite{Ai:2021ghi} toolkit and required to have at least 8 measurements associated to the track fit. Furthermore, at most one expected-but-missing measurement spacepoint (hole) is allowed in each track and track fits are required to have $\chi^2/ndf<3$. Tracks are required to have a $p_{T}$ of at least 500 MeV. No additional requirements on the track transverse and longitudinal impact parameters $d_0$ and $z_0$ are imposed. The combination of these requirements reduces the rate of fake tracks from BIB hits, and their impact on reconstructed jets, to a negligible level.

Hadronic jets are exclusively clustered into two jets using the Valencia algorithm~\cite{Boronat:2014hva} as implemented in \texttt{FastJet}~\cite{Cacciari:2011ma}. The inputs of this algorithm are particle flow objects~\cite{Marshall:2012ry}, which combine track candidates with measurements from the calorimeters using the Pandora toolkit~\cite{Marshall_2012} to improve the jet energy resolution and increase the jet finding efficiency, especially at low jet $p_{T}$. Jets within $|\eta| < 2.44$, with a $p_{T} > 20$~GeV are retained for the further analysis. 

\subsection{Signal and background definition \label{sec:sig-and-back-def}}

To model the signal  
\begin{equation}
  \mu^+ \mu^- \to W^\pm \mu^\mp \nu \to q\, q^{'} \, \mu^\mp \nu
\end{equation}
we generated MC samples with \mgfive~v3.5.6~at tree-level interfaced to Pythia~8.3.13~\cite{Bierlich:2022pfr} for the parton showering and hadronisation. 
%

The main background processes  for both $m_W$ measurement and the later CKM measurements is generic hadrons production 
 \begin{eqnarray}
  \mu^+ \mu^- \to  \textrm{hadrons} \; (+\mu^\mp) + X\,,
\end{eqnarray}
to which there is a significant contribution from the QCD content of the muon, i.e. the quarks and gluons that originate from splitting of the photons that appear in the basic QED muon splitting and give rise to pure QCD scattering processes such as
\begin{eqnarray}
qq (\bar{q}) \to qq (\bar{q}),\;
g q (\bar{q}) \to g q (\bar{q}),\;
q\bar{q} \to gg,\;
\textrm{ and } gg \to gg (q\bar{q}) \;.
\end{eqnarray}

\begin{table}[h]
    \centering
    \begin{tabular}{cc}
    
    \hline
         $qq \to qq$ & 74 fb \\
$q\bar{q} \to q\bar{q}$ & 157 fb \\
$\bar{q}\bar{q} \to \bar{q}\bar{q}$ & 74 fb \\
$q\bar{q} \to gg$ & 2.5 fb \\
         $gq \to gq$ &  529 fb \\
$g\bar{q} \to g\bar{q}$ & 529 fb \\
$gg \to q\bar{q}$ & 30 fb \\
$gg \to gg$ & 824 fb \\
\hline
    \end{tabular}
    \caption{Expected cross-section for jet production form the QCD content of the muon assuming final state partons with $p_T>20\,\gev$. The rates were computed with LePDF interfaced with \mgfive\, via LHAPDF6.
    }
    \label{tab:jet-rates}
\end{table}

\begin{figure}
    \centering
    \includegraphics[width=0.45\linewidth]{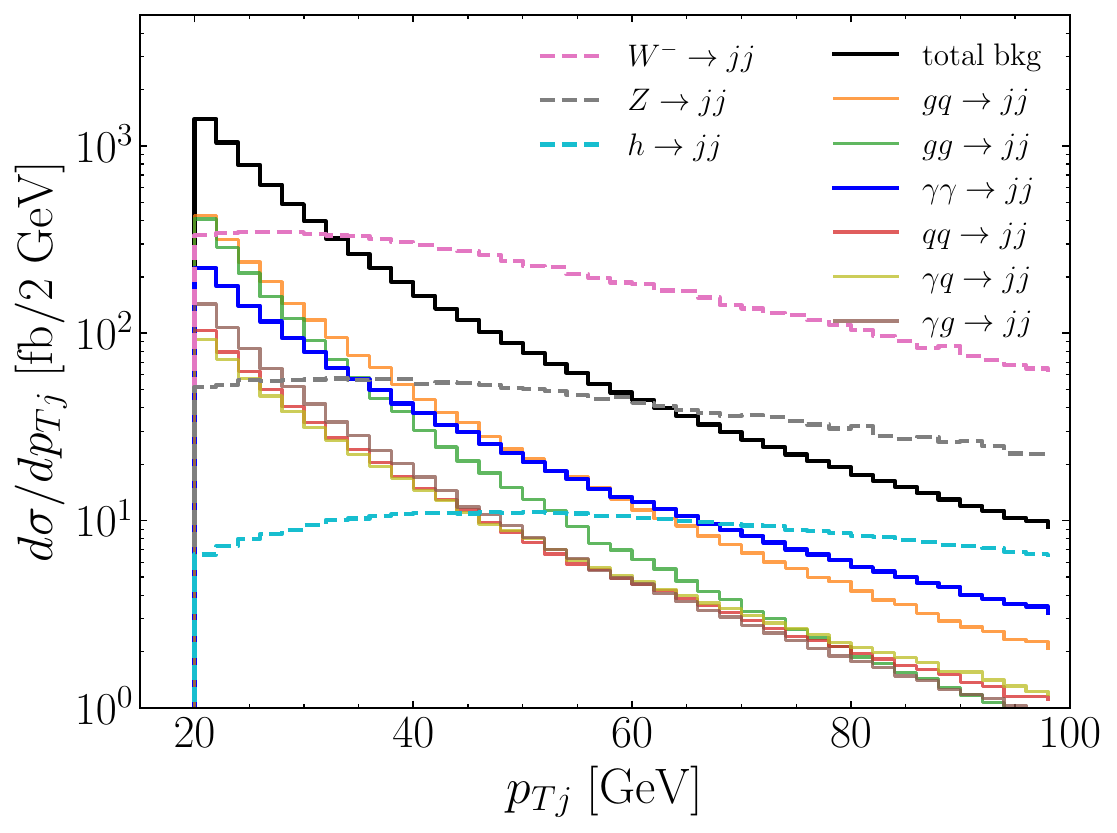}
    \includegraphics[width=0.45\linewidth]{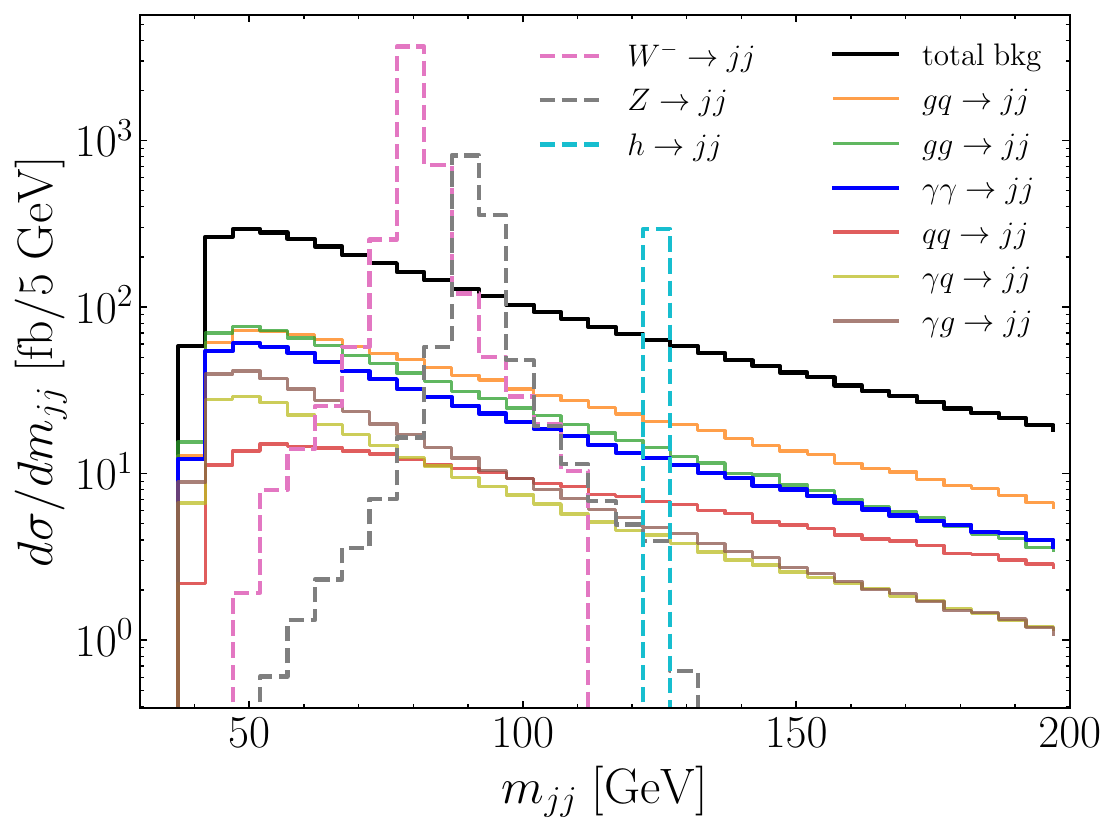}
    \caption{Distribution over inclusive jet $p_T$ and invariant mass of the jets from various QCD, QED and mixed QED-QCD sources of background to the electroweak production of $W$ and $Z$ bosons as computed at LO with \mgfive.}
    \label{fig:QCD-and-QED-jets-pT-and-mass}
\end{figure}

The rates for final state partons at $p_T>20\; \gev$ are given in \Cref{tab:jet-rates}. 
These rates have been computed using LePDF~\cite{Garosi:2023bvq}. In particular,  we implemented an unpolarized version\footnote{Due the chiral nature of the electroweak interaction, the parton content of the muon are intrinsically polarized. But, to have them interface with MG5 in a simpler way, it is a good approximation to treat them as unpolarized at scales not far above $Q\lesssim m_W$.} of LePDF~\cite{Garosi:2023bvq}, interfaced with \mgfive~v3.5.6~\cite{Alwall:2014hca} via LHAPDF6~\cite{Buckley:2014ana}. 
The detailed dependence on jet $p_T$ predicted by our own calculations for each process is shown in  \Cref{fig:QCD-and-QED-jets-pT-and-mass}. 

\Cref{fig:QCD-and-QED-jets-pT-and-mass} shows that in the $m_W$ resonance region there is a significant overabundance of jets from hadronic $W$ over the QCD ones, thus our results are not expected to depend on the precise theoretical knowledge of the shape of the QCD+QED prediction.
A more accurate description of the QCD content of the muon ~\cite{Bauer:2017isx,Han:2020uid,Han:2021kes,Frixione:2023gmf,Garosi:2023bvq}
can be included once  available in public event generators. 
However, we expect our predictions to be sufficient to account for the main effects of the QCD background. 

We  shower the QCD multijets background events with Pythia~8.3.13 before passing them the detector simulation. Unfortunately, the valence structure of the muon implemented in Pythia 8 currently does not support using muon as a beam for QCD partons. Therefore, we use proton/anti-proton beams with muon PDFs to simulate the shower, with setting the option the PartonLevel:ISR = on.

The QCD processes dominate the hadron production rate at small $p_T$. As one considers larger $p_T$ they eventually become smaller than the lower order process 
\begin{eqnarray}
    \gamma \gamma \to hadrons\,,
\end{eqnarray}
which is well known in multi-TeV lepton colliders.
 The transition between hadron production being dominated by QCD and QED processes happens around 50-100~GeV, as also documented in Ref.~\cite{Frixione:2023gmf} and broadly in agreement with our own calculation. This source of hadrons is   considered in our work through a dedicated Monte Carlo sample generated with \mgfive.
 
It is worth nothing that the $Z$ boson decay 
\begin{eqnarray}
    Z \to hadrons
\end{eqnarray} gives rise to jets with kinematics very similar to those from the $W$ boson.  Therefore, the $Z$ boson constitutes a nearly irreducible source of background to our measurements. In the following we take this background into account in detail.

Finally, we also consider
\begin{eqnarray}
    h\to hadrons\;,
\end{eqnarray}
even though it is subdominant compared to $Z\to hadrons$.

\subsection{Templates fitting}

\begin{figure}[h!]
  \center
  \includegraphics[width=0.8\linewidth]{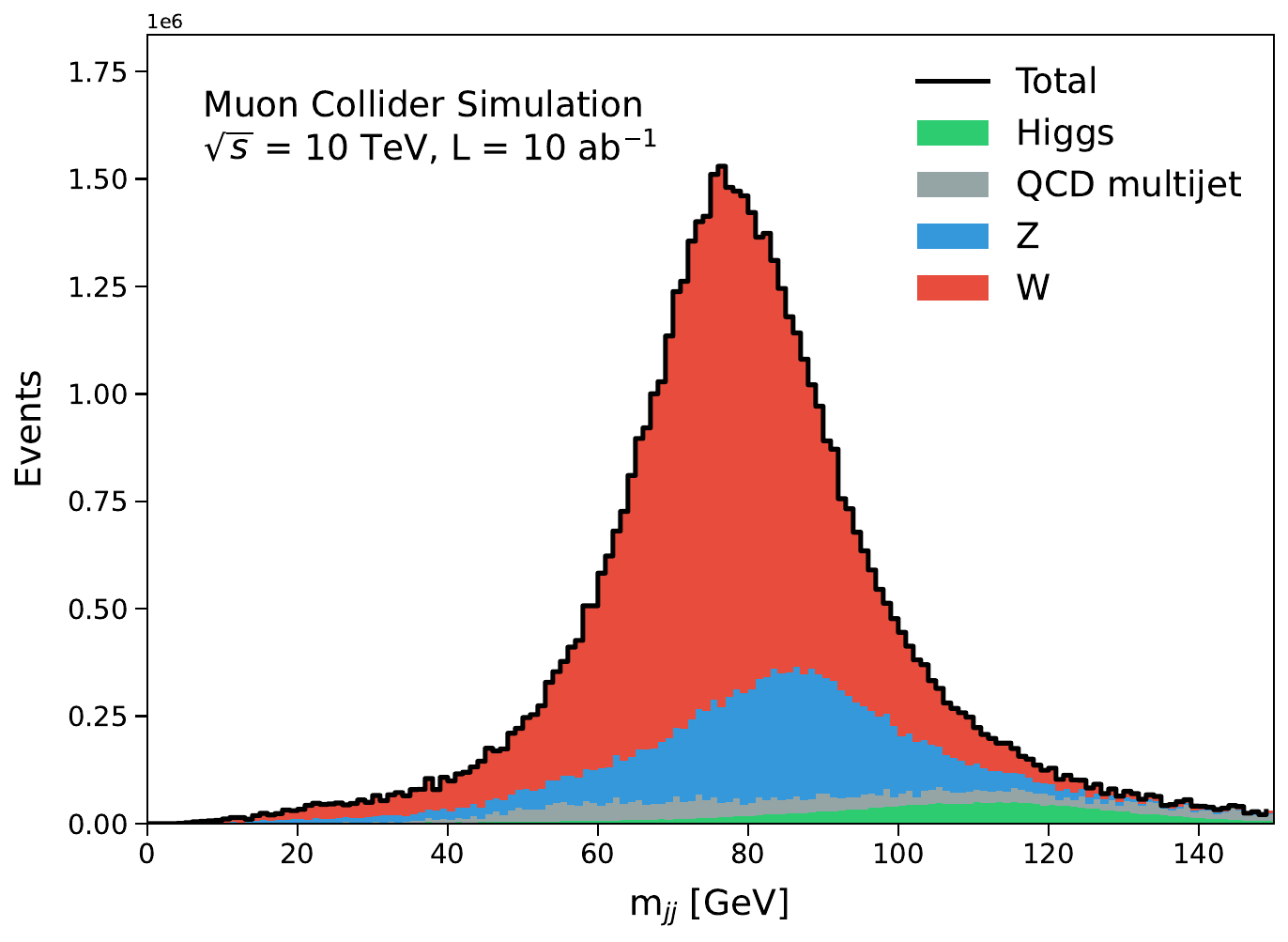}
  \caption{\label{fig:mjj_reco} Distribution of the expected di-jet invariant mass in 10~ab$^{-1}$ of $\sqrt{s}=10$~TeV $\mu^{+}\mu^{-}$ collisions, as recorded by the MAIA detector prior to the in-situ jet energy calibration. The colored histograms represent the contributions of the main processes expected to contribute to this final state. The QCD multi-jet contribution is modeled  as described in Section~\ref{sec:sig-and-back-def}.}
  \end{figure}

The mass of the W boson is extracted from template fits to the invariant mass of the di-jet system. In the absence of prior knowledge of the achievable jet energy scale calibration precision in a future detector, the jet energy scale, $\Delta_{scale}$, is treated as a free parameter and determined simultaneously in the fit. An in situ constraint is provided by the inclusion of hadronic Z boson decays, thereby improving the overall precision of the measurement. The expected final-state distributions (templates) are generated for several values of $m_W$ and the jet energy scale hypothesis, and include the contribution from hadronic Z boson decays with the mass fixed to $m_Z = 91.1876\,\mathrm{GeV}$. These templates are compared to pseudo-data distributions, generated assuming a reference value of $m_W = 80.369\,\mathrm{GeV}$, using a profile likelihood fit. The dominant QCD multi-jet background contribution is estimated from kinematically adjacent sidebands, while other background sources, such as Higgs boson production, are found to have a negligible impact and are therefore neglected.  
To further suppress the QCD multi-jet background, a requirement on the azimuthal separation of the two jets is imposed, $\Delta\phi \leq 2.5$, in addition to a transverse momentum requirement of $p_T^{\text{jet}} > 20\,\mathrm{GeV}$ for each jet. This selection exploits the distinct topology of signal and background processes.
 QCD dijet events, arising predominantly from $2\rightarrow2$ scattering, are produced nearly back-to-back in azimuth, up to soft radiation effects. In contrast, W/Z/h processes involve neutrinos in the final state that carry away significant transverse momentum of order $m_W$, resulting in a reduced azimuthal separation between the reconstructed jets. This simple and robust discriminant provides effective background suppression.  After applying this requirement, a signal-to-background ratio of $S/B \simeq 2.52$ is achieved, with the Z and QCD contributions corresponding to approximately $29.8\%$ and $10.0\%$ of the W signal, respectively.  
  Figure~\ref{fig:mjj_reco} shows the distribution of the expected di-jet invariant mass in $10~\mathrm{ab}^{-1}$ of $\sqrt{s}=10$~TeV $\mu^{+}\mu^{-}$ collisions as recorded by the MAIA detector. The coloured histograms indicate the contributions from the dominant processes in this final state, while the QCD multi-jet component is modelled as described in Section~\ref{sec:sig-and-back-def}.

To reduce the computational cost associated with the generation of the full detector response, the detector effects are parameterised using a smearing function obtained from a fit to the reconstructed di-jet invariant mass resolution, defined at the level of the relative difference between reconstructed and truth quantities, as obtained from the full detector simulation described in Section~\ref{subsec:simreco} .
The templates are subsequently generated using this parameterised response model. This approach allows for a flexible exploration of the measurement sensitivity under different assumptions on future detector and reconstruction performance. 
In the following, the results are presented as a function of the assumed $m_{jj}$ resolution and of the functional form used to describe the detector response. The response is modelled using two resolution smearing functions, considering both an idealised detector scenario, described by a Gaussian smearing, and a more realistic case in which energy losses induce an asymmetric response, modeled by a double-sided Crystal Ball (DSCB) function. The latter parameterization, explicited in equation~\eqref{eq:DSCB}, provides an accurate description of both the Gaussian core and the non-Gaussian tails of the detector response.

\begin{equation} \label{eq:DSCB}
    f_{\mathrm{DSCB}}(m_{jj}) = \begin{cases} \begin{array}{rl} 
        e^{\frac{-t^2}{2}} & \mathrm{if} -\alpha_{\mathrm{low}} \leq t \leq \alpha_{\mathrm{high}} \\
        \frac{e^{-\frac{1}{2}{\alpha_{\mathrm{low}}}^2}}{[\frac{1}{R_{\mathrm{low}}}(R_{\mathrm{low}}-\alpha_{\mathrm{low}}-t)]^{n_{\mathrm{low}}}} & \mathrm{if} \ t \leq -\alpha_{\mathrm{low}} \\
        \frac{e^{-\frac{1}{2}{\alpha_{\mathrm{high}}}^2}}{[\frac{1}{R_{\mathrm{high}}}(R_{\mathrm{high}}-\alpha_{\mathrm{high}}+t)]^{n_{\mathrm{high}}}} & \mathrm{if} \ t \geq \alpha_{\mathrm{high}} \\
        \end{array}
        \end{cases}
\end{equation} 
Where $t = \frac{(m_{jj-\mu_{\mathrm{CB}}})}{\sigma_\mathrm{{CB}}} $, $ R_{\mathrm{low}}=\frac{n_{\mathrm{low}}}{\alpha_{\mathrm{low}}}$, and $ R_{\mathrm{high}}=\frac{n_{\mathrm{high}}}{\alpha_{\mathrm{high}}}$, 
where $\mu_{\mathrm{CB}}$ and $\sigma_{\mathrm{CB}}$ are respectively the mean and the standard deviation of the Gaussian core of the DSCB, $n_{\mathrm{Low}}$ and $n_{\mathrm{High}}$ are the powers of the power-law 
tails on the left side and on the right side of the peak, and finally $\alpha_{\mathrm{Low}}$ and $\alpha_{\mathrm{High}}$ are the parameters describing the connection between the Gaussian core and the tails 
on the left side and on the right side of the peak. In particular, the exponential tails begin for values that differ from the peak by $\alpha \times \sigma $. 
The total number of parameters describing the DSCB shape is therefore six: $\mu_{\mathrm{CB}}$, $\sigma_{\mathrm{CB}}$, $\alpha_{\mathrm{Low}}$, $\alpha_{\mathrm{High}}$, $n_{\mathrm{Low}}$, and $n_{\mathrm{High}}$.

The analysis performs a simultaneous determination of $m_{W}$ and the jet energy scale parameter, $\Delta_{\text{scale}}$, through a two-dimensional template fit in the $(m_W, \Delta_{\text{scale}})$ plane. The likelihood function, which quantifies the compatibility between the pseudo-data and the MC templates, is defined as:
\begin{equation}\label{eq:plh}
{\cal L} \left( \vec n \right | m_{W}, \Delta_{scale}) =
\prod\limits_{i} \text{Poisson} \left(n_{i} | \nu_{i}(m_{W},\Delta_{scale}) \right)\,,
\end{equation}
where $\vec n$ represents the observed distributions in the pseudo-data, and $\nu_i$ is the expectation $\nu_{i}(m_{W},\Delta_{scale}) = W_{i}(m_{W},\Delta_{scale})+Z_{i}(\Delta_{scale})+QCD_{i}$, of $W_{i}$ events from signal, $Z_{i}$ events from Z boson and $QCD_{i}$ QCD multi-jet contributions. %
Uncertainties of the signal and background distributions could be further included as nuisance parameters, distributed according to a Gaussian distribution. However, since the magnitude of the uncertainty on these nuisance parameters at the future experiments is unknown, they were omitted in the final fit.

Nevertheless, the impact of the jet energy resolution on the determination of $m_{W}$ was studied by performing the mass measurement assuming different jet energy resolutions in a range up to 1\%, with a fixed uncertainty of 1\% on the resolution determination, finding a negligible impact on the extracted $m_{W}$ across the whole range.
Dedicated studies have also been performed to evaluate the impact of systematic uncertainties associated with the modeling of the QCD multi-jet background on the extraction of $m_{W}$ and the jet energy scale. Variations in the QCD-background normalisation and shape were incorporated as nuisance parameters and profiled in the two-dimensional likelihood scans. The inclusion of these uncertainties leads to a significant degradation of the measurement precision compared to the statistical-only case, at the level of approximately $40\%$ to $70\%$ worse results depending on the amount of QCD modeling uncertainty.
The impact increases with worsening $m_{jj}$ resolution and constitutes a leading contribution to the overall uncertainty. These results underline the importance of accurate background modeling and improved predictions, the impact of which should be assessed in future studies. 

\begin{figure}
    \centering
    \begin{subfigure}{0.4\linewidth}
    \includegraphics[width=\linewidth]{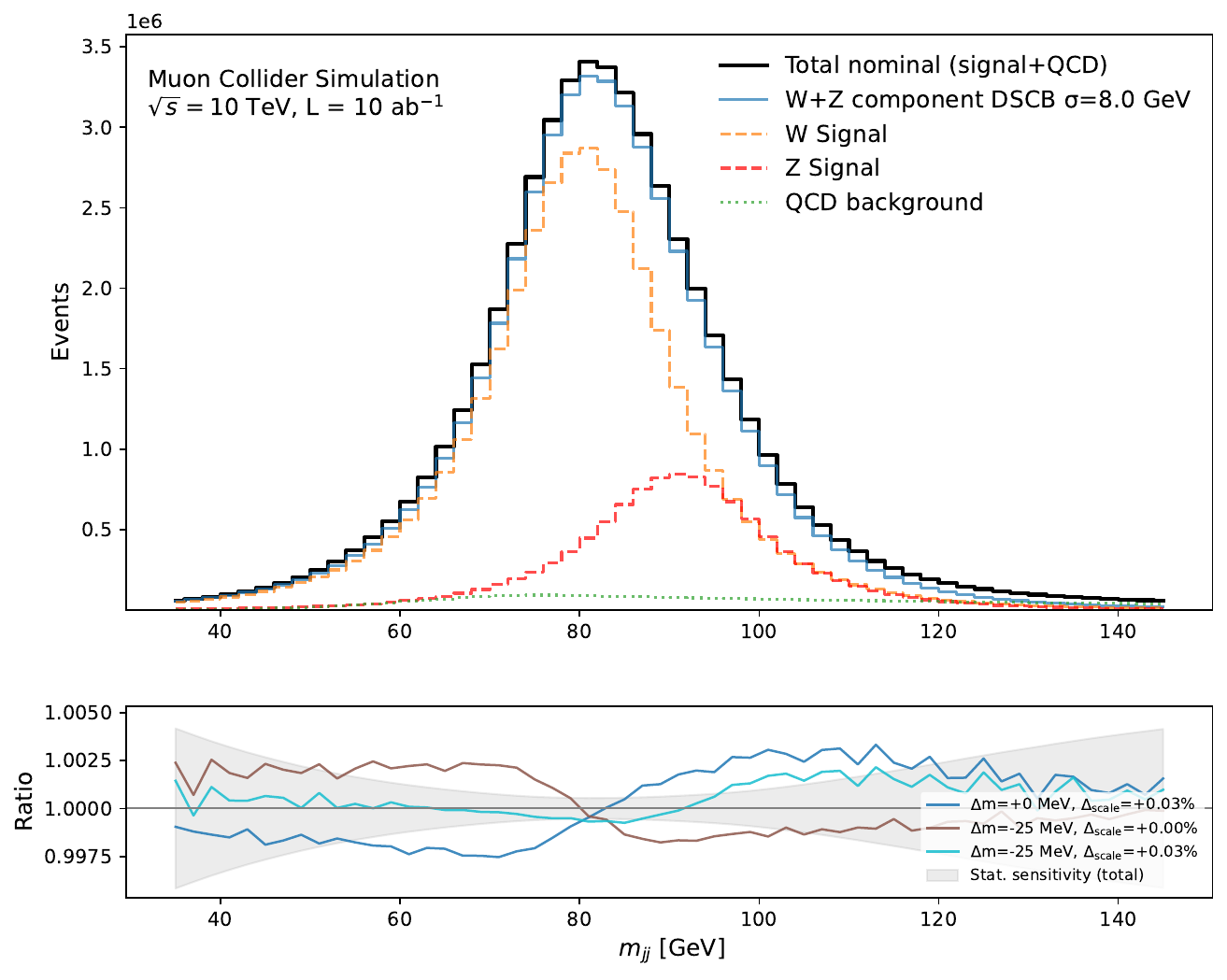}
    \caption{}
    \label{fig:template:distro}
    \end{subfigure}
    \begin{subfigure}{0.48\linewidth}
    \includegraphics[width=\linewidth]{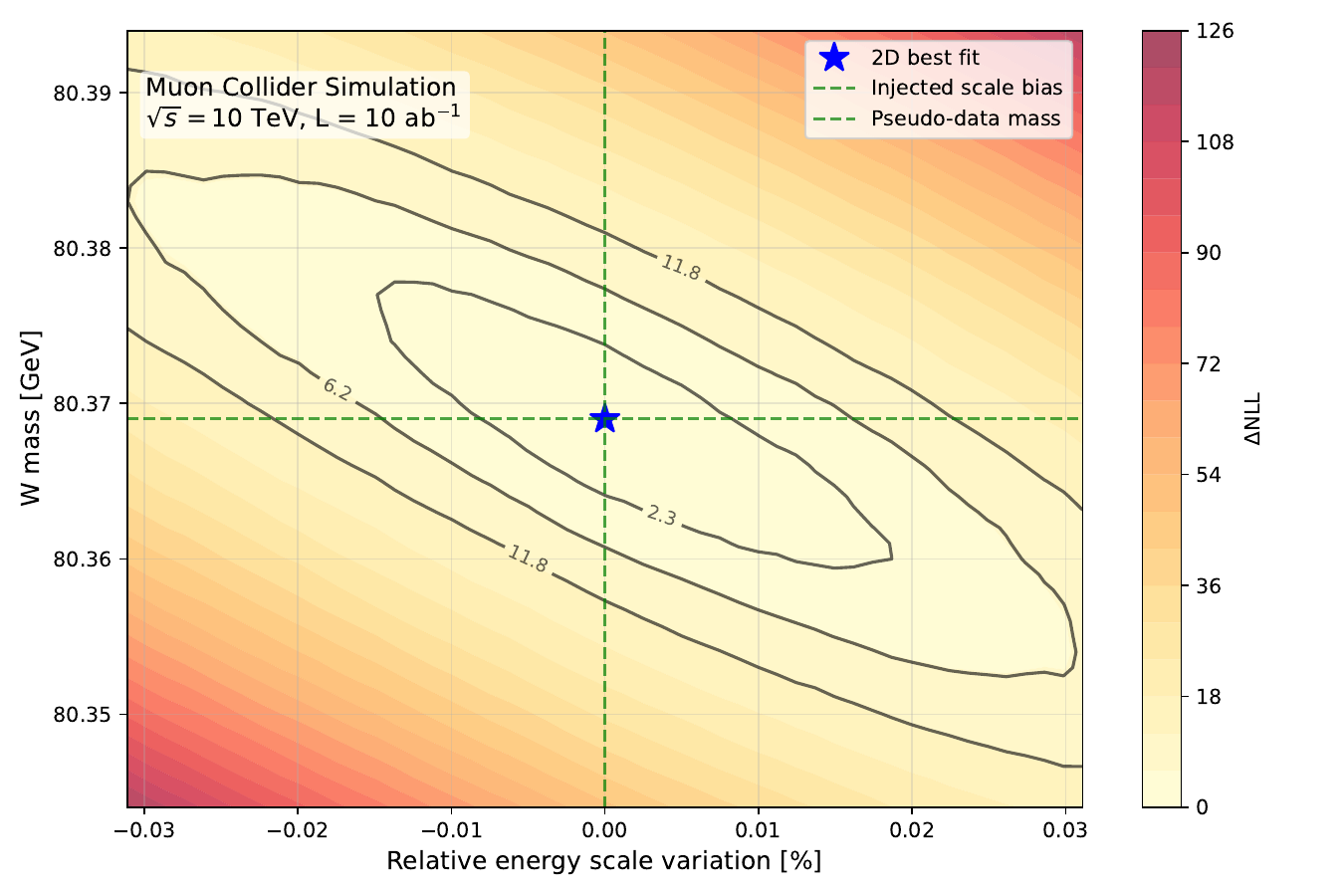}
    \caption{}
    \label{fig:template:2D}
    \end{subfigure}
    \caption{ 
    Example of the generated $m_{jj}$ distribution and the corresponding two-dimensional $(m_W, \Delta_{\text{scale}})$ likelihood scan, shown for a detector response modelled by a double-sided Crystal Ball parameterisation with an $m_{jj}$ resolution of 8~GeV. 
\protect\subref{fig:template:distro} Distribution of $m_{jj}$ for signal and background after event selection. The lower panel illustrates the effect of a $25~\mathrm{MeV}$ variation in $m_W$, corresponding to a jet energy scale shift of $\mathcal{O}(3\times10^{-4})$, compared to the Poisson statistical uncertainty on the pseudo-data (grey band). 
\protect\subref{fig:template:2D} Profile likelihood results expressed in terms of $\Delta\mathrm{NLL}$ as a function of the parameters of interest. The contours indicate the 1-, 2- and 3-$\sigma$ confidence intervals around the best-fit values. }%
    \label{fig:mjj_template}
\end{figure}

The final $W$ boson mass is extracted from two-dimensional likelihood scans in the $(m_W, \Delta_{\text{scale}})$ plane, based on a binned Poisson template fit to the $m_{jj}$ distribution in the range $35\text{--}145\,\mathrm{GeV}$, as defined in Eq.~\eqref{eq:plh}. The scans are performed under several assumptions on the detector response model and the corresponding $m_{jj}$ resolution.  Figure~\ref{fig:template:distro} illustrates a representative example of the signal template obtained for an assumed $m_{jj}$ resolution of 8~GeV. The expected dijet invariant mass distribution is constructed using a detector response modeled by a double-sided Crystal Ball function and is shown for the pseudo-data, as well as separately for the QCD background and the $W$ and $Z$ boson production processes. The lower panel illustrates the effect of $25~\mathrm{MeV}$ variations in $m_W$ and $\Delta_{\text{scale}}$, compared to the Poisson statistical uncertainty on the pseudo-data counts, assuming an integrated luminosity of $10~\mathrm{ab}^{-1}$. A $\Delta_{\text{scale}}$ variation of $25~\mathrm{MeV}$ corresponds to a relative jet energy scale shift of approximately $3\times10^{-4}$. Figure~\ref{fig:template:2D} presents the results of the two-dimensional contours in the plane of the jet energy scale and $m_W$. The contour lines represent the $1-, 2-$ and $3-\sigma$ confidence intervals around the best-fit values. A clear anti-correlation between $m_W$ and the jet energy scale is observed, reflecting the fact that a shift in the overall energy scale can partially mimic a variation of $m_W$ in the reconstructed mass. At fixed resolution, this residual mass–scale degeneracy inflates the statistical sensitivity on $m_W$ by a factor of approximately 1.6–2.0, depending on the assumed resolution.

The results of the template fits are summarised in Table~\ref{tab:hadmW_scan} as a function of the assumed detector response model and corresponding $m_{jj}$ resolution. The quoted uncertainties correspond to one-dimensional profile likelihood intervals extracted from the full two-dimensional $(m_W,\,\Delta_{\text{scale}})$ scan. In this procedure, the complementary parameter of interest is profiled, i.e.\ allowed to vary freely and adjust optimally at each point, thereby accounting for the intrinsic correlation and degeneracy between the $W$ boson mass and the jet energy scale.

The study demonstrates that with the very high statistic collected at muon colliders the hadronic $W$ boson decays can be exploited to extract $m_{W}$ with a statistical precision ranging from approximately $3$ to $6~\mathrm{MeV}$, depending on the detector performance assumptions. The W-mass statistical precision degrades approximately linearly with detector resolution. The upper bound corresponds to an $m_{jj}$ resolution of $10~\mathrm{GeV}$, consistent with the value obtained from a double-sided Crystal Ball fit to the $m_{jj}$ response in the full simulation of the MAIA detector. 
  
Further improvements in reconstruction and calibration techniques are expected to enhance this performance. In addition, the present study is based on a fully inclusive analysis; a more precise determination of $m_{W}$ could be achieved by introducing a categorisation of events according to detector regions or jet kinematics. Despite these conservative assumptions, the projected precision is competitive with existing measurements from LHC experiments.

\begin{table}[t!]
\centering
\begin{tabular}{|c|c|c|c|}
  \hline 
  Template type & $\sigma (m_{jj})$ [GeV] & Expected & Expected relative \tabularnewline
   &  & $\Delta m_{W}$ [MeV] & $\Delta_{scale}$ $[10^{-4}]$ \tabularnewline
  \hline 
  \hline 
  Gaussian & 4 & 2.1 & 0.26 \tabularnewline
  Gaussian & 6 & 3.3 & 0.56 \tabularnewline
  Gaussian & 8 & 4.8 & 0.94 \tabularnewline
  Gaussian & 10 & 5.0 & 0.97\tabularnewline
  \hline 
  DSCB & 4 & 2.7 & 0.40 \tabularnewline
  DSCB & 6 & 3.9 & 0.62 \tabularnewline
  DSCB & 8 & 5.2 & 0.91 \tabularnewline
  DSCB & 10 & 5.3 & 0.82 \tabularnewline
  \hline 
  \end{tabular}
\caption{Summary of the results from the template fits, shown for various assumptions on the data distribution and $m_{jj}$ resolution, for an integrated luminosity of 10~ab$^{-1}$. The results do not include systematic uncertainties. Most notably, as discussed in the text, uncertainties on the QCD multi-jet background are expected to be non negligible, but cannot yet be estimated with the state-of-the-art predictions used in this work.\label{tab:hadmW_scan}}
\end{table}

\section{CKM matrix elements \label{sec:CKM}}
Thanks to the improved detectors that are planned to be in place at future particle colliders, accurate quark flavor tagging is expected to be available and enable new types of analyses. Frontier topics on this matter include plans to measure $h \to s \bar{s}$ with a strangeness tagger~(see for instance  Refs.~\cite{Bedeschi:2022rnj,deBlas:2024ieh,Blekman:2024wyf,Abidi:2025dfw} and references therein), and high precision tagging of heavier $c$ and $b$ flavors, for which already the LHC has shown significant improvements \cite{FTAG-2019-07,FTAG-2023-05,FTAG-2020-08,ParticleNet,CMS-DP-2024-066,CMS-PAS-BTV-22-001}.
The topic has also an important theoretical significance as it insists on foundational aspects of the quark-hadron duality picture.  High precision taggers, especially for quarks such as the strange flavor whose mass is comparable to $\Lambda_{QCD}$,  can potentially shed light on the limitation of our ability to describe hadronization accurately enough to extract theoretically sound information on flavored quantities in general, and  in the case at hand  on the flavor dependence of the $W$ boson couplings. 

Studies on the application of these flavor capabilities are currently motivated mainly from needs in Higgs boson physics, e.g. to try to measure directly each quark Yukawa coupling. 
Given the abundant production of electroweak bosons $W$ and $Z$ that we have reported in \Cref{tab:rates} it is natural to try to extend the application of flavor tagging to measurements such as couplings of the $W$ boson, e.g.  the interactions involved in the decays
\begin{equation}
  W\to q\bar{q}'\, . \label{eq:Wdecay-heavy-flavor}
\end{equation}
While we consider the generic hadronic decays of $W$, with $q, q' = u, d, s, c$, and $b$, the most promising modes at muon colliders, as we will show later in the section, are,
\begin{equation}
W\rightarrow c\, s, \quad W\rightarrow c\, b\, .
\end{equation}
These two modes involve directly CKM matrix elements $V_{cs}$ and $V_{cb}$, which are among the less well-known elements of the CKM matrix to date and might not be improved much in the near future for the reasons explained below.

Measurements of these CKM elements are carried out from flavored hadron decays, which can be experimentally studied at both $e^+e^-$ flavor factories and hadron colliders. A major obstacle to extract short distance couplings such as the CKM matrix elements has to do with the need to have precise predictions for the long-distance part of the decay amplitude. Such long-distance physics effects can be computed on the lattice with non-perturbative methods with some precision. Still, they represent the  largest uncertainty in the extraction of the heavy flavor CKM matrix elements we are interested in. 
The current PDG values of these CKM matrix elements are reported in \Cref{tab:current-CKM}. Broadly speaking these two CKM matrix elements are known today at the level of O(1\%) precision and they are difficult to improve, due to the non-perturbative nature of the leading uncertainty. As a matter of fact, two determinations of $V_{cb}$ are presently in mild tension, which might even be a signal of too optimistic uncertainty estimates\cite{HeavyFlavorAveragingGroupHFLAV:2024ctg,Gambino:2019sif,Bernlochner:2019ldg}.

Given the limitations that can be foreseen in progressing the extraction of $V_{cb}$ and $V_{cs}$ from hadron decays, we explore their measurement from the production rate of $W$ bosons and the observation of their decay into specific flavored final states. 

Our approach is similar to the spirit of~\cite{Marzocca:2024mkc,Liang:2024hox}, which have explored the measurement of $V_{cb}$ and $V_{cs}$ at $e^+e^-$ machines from pair production of $W$ bosons eq.~\eqref{eq:annihilation-pair} considering center of mass energies around 240~GeV.
Here instead we will concentrate on single $W$ production, that, as apparent from \Cref{tab:rates}, is the dominant process for a multi-TeV lepton collider in the range of 10~TeV center of mass energy. { To get a complete picture of the capability of future lepton colliders, we also generalize our study to include all six CKM matrix elements $V_{ij}$ with $ij = ud, us, ub, cd, cs$, and $cb$.} {As already discussed for the mass measurement this process is available at same-sign lepton colliders, thus a $\mu^+\mu^+$ or an $e^-e^-$ machine, up to possible effects due to different beam effects such as beamstrahlung, should do as well as the results we report below for $\mu^+\mu^-$ machines.}

\begin{table}
\centering
\begin{tabular}{|c|c|c|}
  \hline 
   & PDG 2024 \cite{ParticleDataGroup:2024cfk}& rel. unc.\tabularnewline
  \hline 
  \hline 
  $V_{cs}$ & 0.975(6) & $6\cdot10^{-3}$\tabularnewline
  \hline 
  $V_{cb}$ & 0.0408(14)  & $3.4\cdot10^{-2}$\tabularnewline
  \hline 
  \end{tabular}
\caption{Current precision on the CKM matrix elements $V_{cb}$ and  $V_{cs}$. \label{tab:current-CKM}}
\end{table}
 
The signal we consider is the same as for $W$ boson production, with hadronic decay modes,
\begin{eqnarray}
  \mu^+ \mu^- \to W^\pm \mu^\mp \nu \to j\, j \, \mu^\mp \nu\,, 
\end{eqnarray}
where we rely on flavor-tagging of the final state jets to differentiate decay modes with different flavors. {For the description of the generation of the MC samples for the signal and background we refer to \Cref{sec:sig-and-back-def}, with the note that in this section we shall  include detector effects via a parametrized transfer matrix. The transfer matrix is a representative choice from \cite{Bedeschi:2022rnj}. The most crucial performance of the taggers turns out to be its systematic uncertainty, thus  we will reason on the impact of systematic uncertainty and will determine  how  the extraction of the CKM elements degrades as the systematic uncertainty grows.}

In particular, we note that the processes
\begin{eqnarray}
  \mu^+ \mu^- \to W^\pm \mu^\mp \nu \to c\, (b/s) \, \mu^\mp \nu\,, \label{eq:Vcs-Vcb-signal}
\end{eqnarray}
yield around {$6\times10^7$ ($1\times 10^5$)} signal events before considering any acceptance and efficiencies for $cs$ ($cb$) for the standard scenario of 10~ab$^{-1}$ at 10~TeV muon collider. These rates enable an extraction of the CKM elements of interest well below the present uncertainty and possibly competitive with machines operating at the $e^+e^- \to WW$ threshold, where few $10^8$ $W$ boson pairs should be collected. 
Needless to say that the kinematics of the jets would be very different in these two types of colliders and also the detector performances might be significantly different. Therefore, the two studies are synergistic and will offer complementary views on the same measurement. Adding to that, we remark that a completely independent route to measure CKM elements has been proposed for muon colliders exploiting the neutrinos from the $\mu$ beam in a DIS-like experiment~\cite{Marzocca:2025inb}. 

\subsection{Dependence of the CKM matrix elements on systematic tagging uncertainty on the tagging efficiencies}

To estimate the sensitivities on the CKM matrix elements, the number of signal events for $W^\pm\rightarrow i j$ is parametrized as
\begin{equation}
    N_{ij}= N_W\ {\rm Br}_{\rm had}\ \mathcal{A}_W\  \dfrac{|V_{ij}|^2}{\sum_{lm={ud,us,ub,cd,cs,cb}}|V_{lm}|^2},
\end{equation}
where $N_W$ is the number of $W$ boson produced at muon collider, ${\rm Br}_{\rm had}$ is the hadronic branching fraction of the $W$ boson, and $\mathcal{A}_W$ is the acceptance for the hadronic $W$ decay. {We remark this parametrization does not assume the $V$ matrix to be unitary. Additionally, under the assumption that the decay modes   $ud$, $us$, $ub$, $cd$, $cs$, $cb$, in the denominator exhaust the hadronic decay channels of the $W$ boson our analysis does not depend on the uncertainty on the overall production of $W$ bosons. In particular, this renders the results independent of the uncertainty on the total luminosity collected or missing knowledge of the $W$ production rate}
\footnote{In the case of further decay modes, e.g. into new physics states, we remark that our strategy for the extraction of the CKM matrix elements would not suffer from flavor blind new physics contributions. On the contrary, for a flavor dependent new physics hypothesis one would have to consider extra nuisance parameters and simultaneously float the BSM flavor-dependent widths and the CKM elements in a ``unified'' fit, similarly to the spirit of the $m_W$ measurements in presence of new physic discussed in Refs.~\cite{Agashe:2023itp,Agashe:2024owh}.}.
{It is worth remarking that we carry out the discussion without mention of radiative corrections to the $W$ decay, but that these need to be carefully taken into account.}

\begin{table}[t!]
\centering
\renewcommand{\arraystretch}{1.2}
\begin{tabular}{c|cccccc}
\hline\hline
 & $b$ & $s$ & $c$ & $u$ & $d$ & $g$ \\
\hline
$\epsilon_{b}^{\beta}$ & 0.80 & 0.0001 & 0.003 & 0.0005 & 0.0005 & 0.007 \\
$\epsilon_{c}^{\beta}$ & 0.02 & 0.008 & 0.80 & 0.01 & 0.01 & 0.01 \\
$\epsilon_{s}^{\beta}$ & 0.01 & 0.90 & 0.10 & 0.30 & 0.30 & 0.20 \\
\hline\hline 
\end{tabular}
\caption{Jet-flavor taggers working points~\cite{Bedeschi:2022rnj,Marzocca:2024mkc}: the probabilities $\epsilon_{q}^{\beta}$ to tag a jet of true flavor $\beta$ as a $q$-jet, with $q=\{b,c,s\}$ and $\beta=\{b,s,c,u,d,g\}$. For each column a fourth category with probability $1-\epsilon^\beta_b - \epsilon^\beta_c -\epsilon^\beta_s $ accounts for the case that a true flavor $\beta$ is not tagged. Rather than re-optimising the choices of working points for this work, we opted to use the same combination used in other works. In the future, when detailed expectations for the flavour tagging performance at muon colliders will become available, this study should be repeated and optimised to maximise the sensitivity.
}
\label{tab:jetflavor_wp}
\end{table}

{ For the extraction of the CKM elements $V_{ij}$, we define ten mutually exclusive and exhaustive tagging bins according to the joint number of $b$-tagged, $c$-tagged and $s$-tagged jets in each event,
\[(n_b, n_c, n_s),
\]
where the three entries correspond to the number of $b$-tagged jets, $c$-tagged jets, and $s$-tagged jets, respectively. As we will show later, the most promising improvements that can be made at muon colliders are those on the precisions of $V_{cs}$ and $V_{cb}$, in which cases the dominant constraining power comes from the $(0,1,1)$ bin and $(1,1,0)$ bin, respectively.}

This binning captures the full combinatorial flavor information relevant to the $W$ boson decay while keeping the number of bins manageable. Mis-tagging between $c$, $s$, $b$, and light-flavor jets is included through the efficiency matrix $\epsilon^\beta_{q}$, which maps the true flavor $\beta$ to the detected flavor $q$ according to the tagger. The matrix that we take as reference point is displayed in \Cref{tab:jetflavor_wp} and coincides with the choice of Ref.~\cite{Marzocca:2024mkc} for FCC-ee studies~\cite{Bedeschi:2022rnj}. This choice is just an educated guess at this point of the development of the detectors for high energy lepton colliders, but it should serve as a reasonable reference. Using this matrix and the rates for the production of each flavor combination we can populate each of the ten bins and evaluate the uncertainties on its number of counts.

Each bin is modeled as an independent Poisson process,
\begin{equation}
P(N_i^{\rm obs}\,|\,N_i(\theta,\epsilon (\nu))) 
= \frac{ \left( N_i(\theta, \epsilon(\nu))\right)^{N_i^{\rm obs}} }{N_i^{\rm obs}!} e^{-N_i(\theta,\epsilon(\nu))}\;\;,
\end{equation}
{where $N_i$ and $N_i^{\rm obs}$ are the number of counts in the $i$-th bin predicted and observed, respectively.}  

In each bin we include background contributions on top of $W$ decays. The background is simulated at the parton level, as described in Sec.~\ref{sec:sig-and-back-def}. We note that the QCD+QED dijet and the hadronic $Z$ backgrounds have very mild impact on the sensitivities on the CKM matrix elements. Therefore, we only impose mild cuts for background rejection:
\begin{equation}
    p_{Tj} > 20~{\rm GeV,}\quad 10^\circ < \theta_j<170^\circ, \quad 60~{\rm GeV} < m_{jj} < 100~{\rm GeV}.
\end{equation}

The parameter of interest, that is the CKM element under study $ V_{ij}$, is denoted as $\theta$. 
{The efficiency matrix for tagging is denoted as $\epsilon(\nu)$ to indicate that it is a function of some nuisance parameters $\nu$, with each of the matrix entries in principle affected by different bits of detector effects and theory uncertainties.}

{At present there is no available modeling of the detailed sources of systematics on the taggers. We highlight that this is a field much in need of progress, as theory understanding of taggers beyond lowest order approximation of perturbation theory is necessary to ensure systematics at a level suitable for the target precision on CKM elements. Similarly, detector studies are needed to explore up to what point the relevant systematics can be pushed for each flavor. For lack of a more motivated pattern, we assume that the nuisance parameters will result in a common relative uncertainty for all the entries of the $\epsilon$ matrix. We denote this relative uncertainty as $\delta_\epsilon$, and we assume that it  encapsulates both theory and experimental systematics. Bearing these assumptions in mind, we assign to all entries of the matrix  }
a Gaussian prior centered on its nominal value $\bar{\epsilon}_{ j}$ and systematic {relative uncertainty} $\delta_\epsilon=\delta\epsilon/\epsilon$,
\begin{equation}
\pi(\epsilon_{ j})=\exp\!\left[-\frac{(\epsilon_{ j}-\bar{\epsilon}_{ j})^2}{2\,(\delta_\epsilon\bar{\epsilon}_{ j})^2}\right].
\end{equation}
The full likelihood is then
\begin{equation}
\mathcal{L}(\theta,\epsilon) =
\prod_{i} P(N_i^{\rm obs}\,|\,N_i(\theta,\epsilon))
\prod_{j} \pi(\epsilon_j).
\end{equation}
The final constraint on $\theta$ is obtained by profiling over 
$\epsilon$, that is,
\begin{equation}
\mathcal{L}_{\rm prof}(\theta) =
\max_{\epsilon} \mathcal{L}(\theta,\epsilon),
\end{equation}
and extracting the 1-$\sigma$ uncertainty from the curvature of $-2\ln \mathcal{L}_{\rm prof}$ or a constraint on multiple parameters by suitable slicing of the likelihood.

\begin{figure}[t!]
\centering
\includegraphics[width=0.48\textwidth]{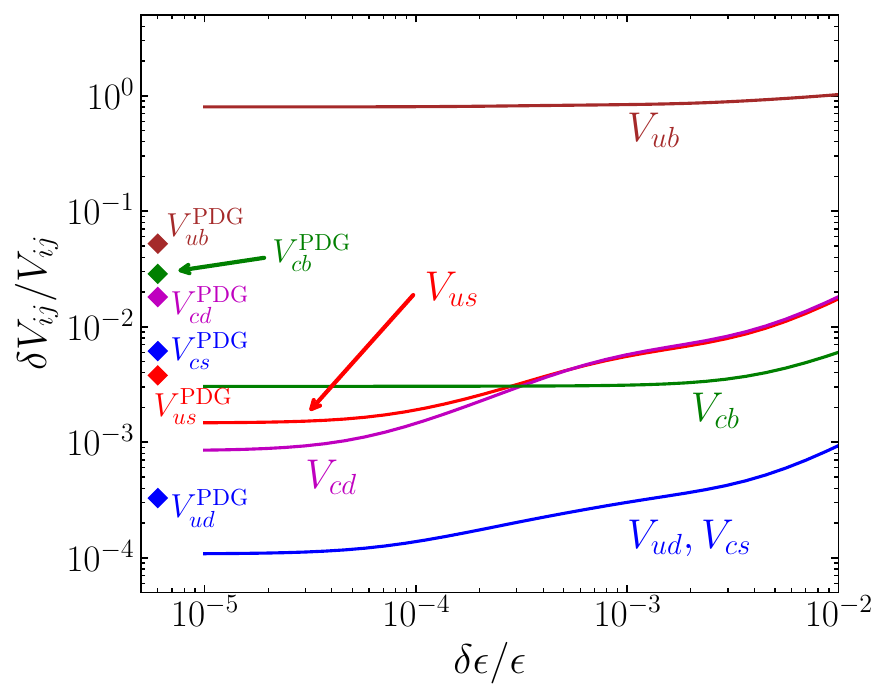}
\includegraphics[width=0.48\textwidth]{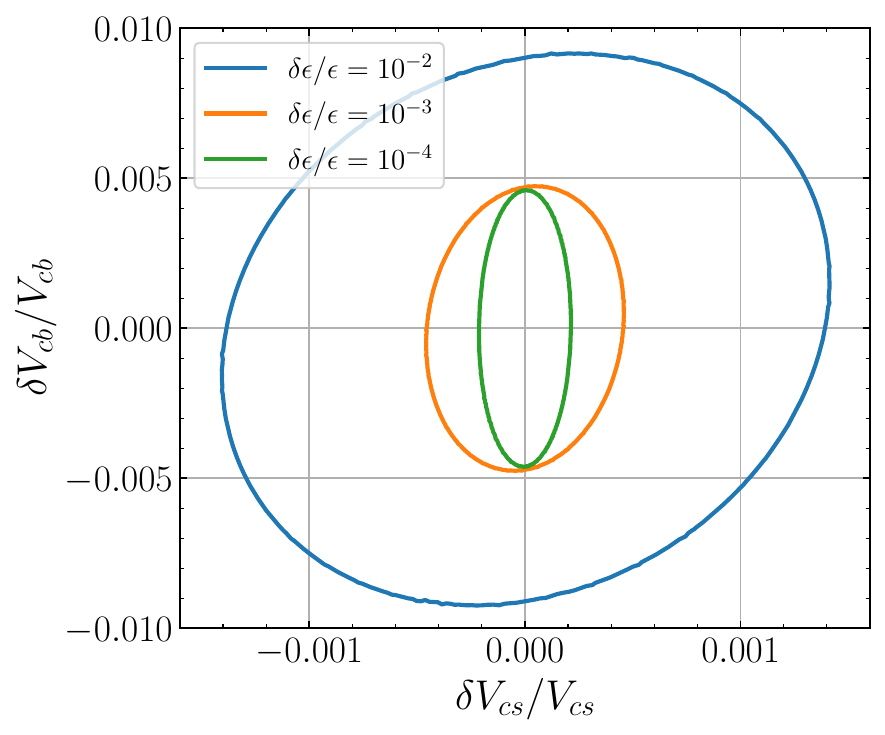}
\caption{Left: Projected 1-$\sigma$ constraints on $|V_{ij}|$ from single-$W$ as a function of the flavor tagger systematics at 10~TeV muon collider for 10~ab$^{-1}$ luminosity. Right: Projected 1-$\sigma$ contours for simultaneous extraction of $V_{cs}$ and $V_{cb}$ from event counts binned according to $(n_c, n_s, n_b)$. 
}
\label{fig:ckm_bounds}
\end{figure}

{The left panel of \Cref{fig:ckm_bounds} illustrates the 1-$\sigma$ constraints on the 1D fits of the CKM matrix element $V_{ij}$, as functions of the systematic uncertainties on the flavor-tagging efficiencies. The current PDG uncertainties~\cite{ParticleDataGroup:2024cfk} are marked on the left.}

{We note that the precision on $V_{cs}$ and $V_{cb}$ can be significantly improved with respect to the current determinations limited by knowledge of hadronic information from the lattice.}
 The sensitivity on $V_{ub}$ is extremely weak due to its small value and the large background from $W\rightarrow ud$ and $W\rightarrow cs$ through mis-tagging. Even for very optimistic small systematic uncertainty we find that with the 10/ab statistics of the 10~TeV muon collider this coupling is hardly improved. While attaining larger statistics can help to reduce the uncertainty, the jump in luminosity to get to an interesting level is more than one order of magnitude compared to the present baseline luminosity of the project.
The precision on $V_{us}$, $V_{cd}$, and $V_{ud}$ can improve with respect to the present determination if the systematic uncertainties on the flavor-tagging efficiencies will be better than $10^{-3}$, setting a clear target for the development of flavor-tagging algorithms.

In the right panel of \Cref{fig:ckm_bounds} we show an example of simultaneous extraction of $V_{cs}$ and $V_{cb}$. We find that the correlation between $V_{cs}$ and $V_{cb}$ is rather mild, and becomes negligible for systematic uncertainties less than $10^{-3}$.

\section{Conclusion \label{sec:conclusion}}
 Our study shows that a high‑luminosity muon collider can determine the mass of the $W$ boson with an accuracy of a few MeV, the exact figure depending on the ultimate performance of the detector on jet energy determination.  Such a measurement would allow to perform   detector calibration and in-depth tests of the reconstruction performance, particularly for the reconstruction of hadronic jets and the tagging of quark flavors.

Moreover, the copious production of $W$ bosons at a muon collider offers a unique opportunity to extract the elements of the CKM matrix. Similarly to attempts at the LHC, see e.g.~\cite{ATLAS:2026gkw}, and prospected extractions at future $e^+e^-$ machines~\cite{Marzocca:2024mkc,Liang:2024hox}, we studied the extraction of CKM elements from rate measurements involving $W$ boson production and decays. 
By exploiting the high‑statistics sample of hadronic decays, the CKM matrix elements can be measured with systematic uncertainties that improve as detector technologies advance. These improvements will be independent from advances in the measurements from hadron decays at heavy flavor factories and theory improvements in lattice QCD. Thus the projected sensitivity to CKM matrix elements will be complementary to these other more traditional extractions of the CKM elements.  Furthermore the method we studied can be combined with the measurement proposed at muon colliders exploiting secondary $\nu$ beams~\cite{Marzocca:2025inb}.

To fully realize these physics goals, further work is needed in key areas such as the development of superior jet reconstruction and energy determination, the development of dedicated flavour‑tagging algorithms suited to the muon‑collider environment, and the development of more refined background and signal predictions to match the required precision for the measurements we are prospecting.  

In closing, we remark that the required developments are  synergetic with the needs for studies based on tags of Higgs boson decay products and absolute rates predictions, e.g. $h\to s\bar{s}$. Further utility of the development pointed by our study can be found noting that the experimental strategies discussed here for $\mu^+\mu^-$ 
can broadly 
be applied to any high‑energy lepton machine that accesses the  
$\gamma W \to W $ production channel.  
In particular, both $e^-e^-$ and $\mu^+\mu^+$ colliders would inherit the same statistical reach 
and the same detector requirements for hadronic energy resolution and flavour tagging, making 
the precision program essentially transferable among these configurations.

\section*{Acknowledgements}
We thank Dario Buttazzo, Roberto Di Nardo, Stefano Frixione, David Marzocca, Davide Pagani, Michele Tammaro, Andrea Wulzer, Yang Ma for useful discussions and the International Muon Collider Collaboration (IMCC) for fostering this
effort. We also thank Massimo Casarsa, Davide Zuliani, and Sarah Heim for comments on the manuscript. RF and XW are supported in part by the European Union - Next Generation EU   Missione 4 - Componente 2 - Investimento 1.1 CUP: I53D23000950006 through the MUR PRIN2022 Grant n.202289JEW4. This work was supported by the EU HORIZON Research and Innovation Actions under the grant agreement number 101094300. Funded by the European Union (EU). Views and opinions expressed are however those of the author(s) only and do not necessarily reflect those of the EU or European Research Executive Agency (REA). Neither the EU nor the REA can be held responsible for them. This work has benefited from computing services provided by the German National Analysis Facility (NAF). RF, FM, XW thank the  Galileo Galilei Institute of INFN for support during the workshop ``Exploring the energy frontier with muon beams'' where part of this work has been carried out.

\bibliography{SciPost_Example_BiBTeX_File.bib}


\end{document}